\documentclass[fleqn,usenatbib]{mnras}
\usepackage{graphicx}
\usepackage{url}
\usepackage{longtable}
\usepackage{float}
\usepackage{subfigure}
\usepackage{newtxtext,newtxmath}
\usepackage[T1]{fontenc}
\usepackage{amsmath}
\usepackage{soul}
\usepackage{lscape}
\usepackage{multirow}

\DeclareRobustCommand{\VAN}[3]{#2} \let\VANthebibliography\thebibliography
\def\thebibliography{\DeclareRobustCommand{\VAN}[3]{##3} \VANthebibliography}

\title[Gas counterparts of YSO associations]{Gas content and evolution of a sample of YSO associations at $d \lesssim$ 3.5 kpc from the Sun}
\author[Zhou, Li \& Chen]{
Ji-Xuan Zhou,$^{1}$$^,$\ $^{2}$ 
Guang-Xing Li,$^{1}$\thanks{gxli@ynu.edu.cn} 
Bing-Qiu Chen$^{1}$\thanks{bchen@ynu.edu.cn} \\
$^{1}$ South-Western Institute for Astronomy Research, Yunnan University, Chenggong District, Kunming 650091, P.\,R. China\\
$^{2}$ School of Physics and astronomy, Cardiff University, Queen's Buildings, The Parade, Cardiff, CF24 3AA, UK}

\begin{document}
\label{firstpage}
%\pagerange{\pageref{firstpage}--\pageref{lastpage}}

\maketitle
\begin{abstract}

    Young Stellar Objects (YSO) are newly formed stars from molecular clouds. They stay close to where they were born and serve as good tracers to study gas and star formation. During cloud evolution, young massive stars can disrupt the surrounding gas through stellar feedback, changing the gas distribution. We study the distribution of the gas around a sample of YSO associations located at $d \lesssim 3.5 \;\rm
    kpc$ from the Sun by comparing the location and morphology between $^{12}$CO (J = 1$-$0) emission, Planck 870 $\mu$m maps and YSO associations. Based on the spatial distribution of the gas compared to that of the YSOs, we classify the YSO associations into six types: direct, close, bubble, complex, diffuse, and clean associations. The complex associations are large structures consisting of both gas-rich and gas-poor segments. We study the velocity dispersion-size relation toward different association types. From the ratio between different types, we estimate a feedback time of $\approx$ 1.7 Myr in the solar neighborhood. The sample sets a solid foundation to explore the relationship between interstellar medium evolution, star formation, and Galaxy structure.

\end{abstract} 

\begin{keywords} % commenting results in missing thanks notes!
    Galaxies: ISM  -- galaxies: star clusters -- ISM: structure -- ISM: clouds  -- Stars: formation 
\end{keywords}
% \noindent {\bf Key words:} 
% {\bf Try to use gas-rich, gas poor when necessary!}

\section{Introduction}

Stars form in molecular clouds, whose collapse is driven by gravity
 \citep{2007ARA&A..45..565M} yet affected by other processes like turbulence
 \citep{1981MNRAS.194..809L, 2000prpl.conf....3V, 2004RvMP...76..125M}, magnetic
 field \citep{2014prpl.conf..101L} and
 stellar feedback \citep{2013A&A...550A..49K,2014prpl.conf..243K}. Instead of being alone, stars form in a clustered way
 \citep{1995MNRAS.272..213L, 1983ARA&A..21..343A, 2003ARA&A..41...57L}. Stars born in the same cloud would retain some properties from the parent molecular clouds. At a smaller scale, the spatial distribution of young stars appears to be structured,  with clusters being associated with the high-density regions in clouds  \citep{1993prpl.conf..245L, 1993prpl.conf..429Z}.

Young stellar objects (YSOs) are young stars born from molecular clouds. 
 Being young \citep{2014A&A...561A..54R,
 2015A&A...576A..52R, 2008ARA&A..46..339W, 2015ApJS..220...11D}, most YSOs are still associated with their parent molecular gas, making them good tracers to the molecular clouds. 
 In observations, YSOs have distinct observation signatures that span across multiple wavelengths, e.g. infrared excess, ultraviolet lines, X-ray emissions \citep{1983ApJ...274..698W, 1987ApJ...312..788A, 2004ApJS..154..363A, 2004ApJS..154..367M, 2008ApJ...688.1142K, 2010AJ....139.1338F,2011ApJ...727...62R, 2013ApJ...774..101R, 1999ARA&A..37..363F, 2004A&ARv..12...71G}.
 In recent decades,  sensitive surveys with high resolutions and large coverages from X-ray to infrared bands have made it possible to identify more YSOs and provide good chances to study their properties and distributions systematically \citep{2004ApJS..154..363A, 2004ApJS..154..367M, 2008ApJ...688.1142K, 2010AJ....139.1338F,2011ApJ...727...62R, 2013ApJ...774..101R, 1999ARA&A..37..363F}.

 In recent years, accurate parallax measurements and photometric measurements of young stars provided by Gaia satellite \citep{2018A&A...616A...1G, 2023A&A...674A...1G}, have revealed a picture of a dynamically evolving galactic disk. 
 \citet{2018A&A...620A.172Z} provided a study of the star formation region in the solar neighborhood using the young stars from Gaia DR2. 
 \citet{2023ApJS..265...59Z}derived the distance of 63 molecular clouds by matching them with over 3000 YSOCs (YSO cluster).
 Besides the location study, 
  \citet{2021A&A...647A..91G} studied the relation between the mean velocity of YSOs and that of the gas in the Orion molecular cloud, and found a good consistency, proving that YSOs are robust tracers of the cloud kinematics. 
 \citet{2022ApJ...934....7H} use the 6D measurements of locations and velocities of young stars in Orion, Ophiuchus, Perseus, and Taurus star-forming region and studied the turbulence by building the velocity structure function. Using kinematic information obtained towards young stars, \citet{2022Natur.601..334Z} studied the structure and expansion kinematics and star formation of gas surrounding the Local Bubble.
 As the young stars evolve in the association, they start to disrupt the ISM in nearby clouds, which might lead to a complete removal of the gas \citep{2010ApJ...710L.142F, 2011piim.book.....D, 2022arXiv220309570C}. During this process, the changing gravitational potential leads to the final dissolution of the associations \citep{2016A&A...587A..53K}. By studying the distribution of young stars with respect to that of the gas,  we can obtain a systematic view of cloud evolution and gas removal.

 %Yet when a cloud is dispersed by processes such as the stellar feedback \citep{2010ApJ...710L.142F, 2011piim.book.....D, 2022arXiv220309570C}, the YSOs will retain their circumstellar material and remain identifiable. 
%  From harboring young stars, the molecular gas go through a lot of star formation activities and stellar feedback and will be destructed and depleted \citep{2009ApJS..184....1K, 2017A&A...601A.146C}.
%   Studying the
%  relations between YSO association and the surrounding gas should lead to the understanding of the effects of
%  stellar feedback on the dispersal of molecular clouds, which is a critical part of our understanding of the star formation process. 

We are now working on the ISM-6D program, in which we combine the CO map \citep{2001ApJ...547..792D}, dust map from Planck \citep{2014A&A...571A...1P}, and Gaia astrometry measurements to study the structure and full kinematics of ISM (interstellar medium) \citep{2016A&A...595A...1G}. The 2D velocities (proper motion in the $l$ and $b$ directions) and locations of YSOs are from Gaia astrometry measurements. The CO data from \cite{2001ApJ...547..792D} provides the radial velocity as the third velocity component. In the program, we use the YSO association to trace molecular clouds. The YSO associations are groups of YSOs clustered in both the spatial and the kinematic space.
In \citet{2022MNRAS.513..638Z} we constructed
a sample of 150 YSO associations. In this paper, we measure the distribution of gas in associations and study the relationship between the spatial distribution of the gas and the YSOs. We classify the YSO associations based on the similarity between the spatial distribution of the gas and that of the YSOs. We reveal a diverse range of ways through which they relate, hinting at a picture that as star formation precedes, the molecular gas is gradually removed by stellar feedback. We link YSO association types with their location in the Galaxy and in the velocity dispersion-size plane, which is a first step in
constructing a sample of YSO-giant molecular cloud (YSO-GMC) complex with complete kinematic information.

\section{Data \& Method}

\subsection{Data}
 
\subsubsection{YSO associations}
We start with a YSO association catalog produced in our recent work \citep{2022MNRAS.513..638Z}.  We use a YSO sample from \citet{2016MNRAS.458.3479M}, which contains 133980 class I/II YSOs using the Support Vector Machine method. After cross-matching the YSO sample with the Gaia DR2 data set, we applied \texttt{Dendrogram} method \citep{2008ApJ...679.1338R} to the YSO density map in $l$, $b$, log($d$) space, which gives us 150 structures, including less compact branch structures and their high-density substructures. These spatially and kinematically clustered YSOs are called YSO associations. Each association contains the following parameters:   location $l$, $b$, $d$, $X$, $Y$ and $Z$, mean proper motion pm$_{l}$ and pm$_{b}$, 2D velocity dispersion, as well as  size \citet{2022MNRAS.513..638Z}.

\subsubsection{Gas tracers}
To study the gas content of the YSO associations, we use a map of
$^{12}$CO ($J$ = 1$-$0) line presented in \citet{2001ApJ...547..792D}, and a map
of 870$\mu$m emission produced by the Planck satellite
\citep{2014A&A...571A...1P}, both of which directly trace the cold molecular
gas.

The $^{12}$CO ($J$ = 1$-$0) rotational line is a good tracer of the cold,
molecular gas. CO data from \cite{2001ApJ...547..792D} is a composite CO map containing data from mostly the CfA 1.2 m telescope. The survey covers the Galactic plane at resolutions ranging from 9$'$ to 18$'$. The map is velocity-resolved with a resolution of 1.3 $\rm km\;s^{-1}$. This enables
us to distinguish different clouds along the line of sight and measure the
radial velocities of the clouds of interest. Although there are other newer CO surveys with higher resolutions,  the complete coverage of data from   \citet{2001ApJ...547..792D} makes it most suitable for our study. 

The Planck dust map we use is taken at 857 GHz \citep{2014A&A...571A...1P}.
The Planck observations contain nine frequency bands, including 30, 44, 70, 100,
143, 217, 353, 545, and 857 GHz. Maps at lower frequencies are used to study
the CMB fluctuations \citep{2014A&A...571A...1P}. The 870$\mu$m map from the
Planck satellite is dominated by emissions from cold dust in molecular clouds,
making it a good tracer for clouds. Compared to the CO map, the Planck dust map
has a slightly higher resolution of 5$'$. One major benefit of the map is its
complete, all-sky coverage, making it particularly useful for high-latitude
regions where the CO map is incomplete.

\subsection{Gas counterparts of YSO associations}   \label{gas counterpart}                                                     
% To study how gas is distributed around the YSOs in the association. To achieve this, we plot the member YSOs on both the CO map and the Planck map, both can trace the distribution of the gas. 

%find the multiple cloud components from CO map along the direction of the YSO association. We plot the member YSOs in the association on the multiple CO component map and choose the best-matched cloud.  We also plot member YSOs on the Planck map, and then compare the spatial distributions of YSOs and the nearby gas in $l$ and $b$ plane. 

We aim to study how gas is distributed around the YSOs in the associations. To achieve this, we plot the member YSOs on both the CO map and the Planck map, which can trace the gas. The CO map from \cite{2001ApJ...547..792D}  contains velocity information. At a given location, different clouds appear as different velocity components. This velocity information is useful as the YSO associations are likely to be associated with one of these velocity components, and we aim to find the appropriate component at a similar location on the sky plane. To achieve this, 
towards the CO data cube, first, we integrate the CO emission over a box-like area in the position-position space. The area is the footprints of the YSO associations\footnote{The chosen region is defined as
[${\rm min}(l_i)-3^{\circ}:{\rm max}(l_i)+3^{\circ}$] and [${\rm min}(b_i)-3^{\circ}:{\rm max}(b_i)+3^{\circ}$],
where $b_i$ and $l_i$ represent the galactic latitude and longitude of member
YSOs in association $i$, respectively.}. 
The integrated emission is then plotted along the radial velocity axis, producing the line profile and showing gas distribution at different velocities.  

The resulting CO line profiles would contain several Gaussian-like components if there are several different clouds along the line of sight, and the corresponding cloud should appear as one of these components. In our second step, we perform decompositions to the integrated line profile. The decomposition is performed using the \texttt{Gaussian Mixture} algorithm in
\texttt{python} \citep{lindsay1995mixture, pedregosa2011scikit}, where we set the maximum Gaussian component number $N$ and input the line profile. The algorithm fits the line profile using 1 to $N$ Gaussian components, where the optimal decomposition output is selected using the  Akaike Information Criterion (AIC), which estimates the relative error between the data and the fitting model. The one with the smallest AIC value is the one we chose.

Based on the decomposition results, we produce maps of the CO emissions into different velocity components and select the corresponding cloud by comparing the spatial distribution of YSOs with the morphology of the gas or dust emission (details and examples are available in Appendix \ref{appendixa}).

The Planck dust map contains no velocity information. We plot the member YSOs in each association on the dust map directly and compare the dust distribution and morphology with the member YSOs in the $l$-$b$ space through visual inspections.

%Not all the associations have detectable gas counterparts traced by CO or dust. In Fig. \ref{matching example} and Fig. \ref{class_example}, we present examples illustrating both cases. When the YSO associations are well associated with gas(like gas-rich associations), the spatial distribution of member YSOs resembles the distribution of the gas; when the gas and YSO associations are not associated (like gas-poor associations), no clear gas counterparts are seen. 

\begin{figure*}
\begin{center}
\includegraphics[scale=0.35]{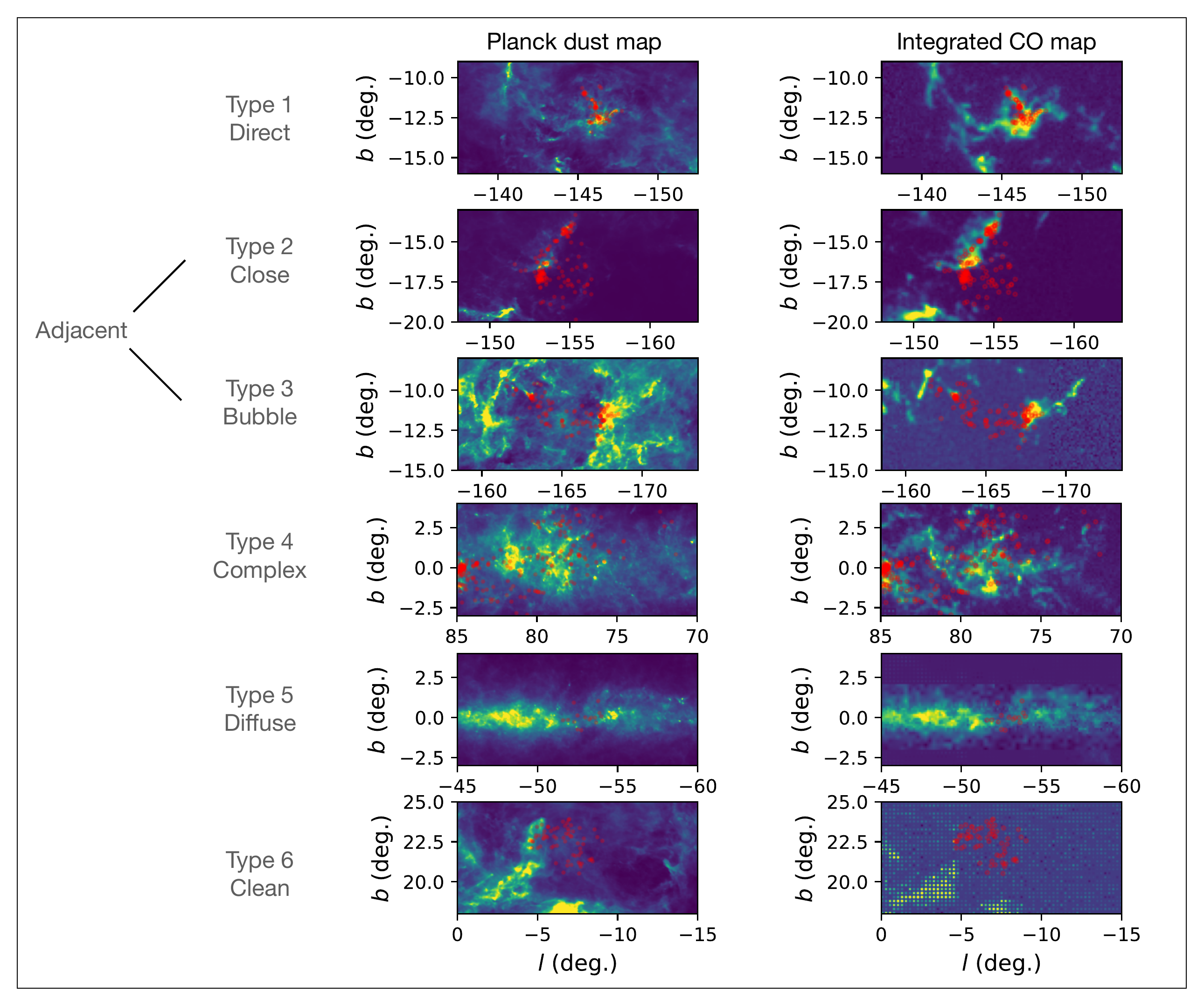}  
\caption{{\bf Examples of different types of YSO associations.} Red dots in all
the subplots represent the member YSOs in the corresponding YSO associations. The left panels are the member YSOs plotted against the dust map form
\citet{2014A&A...571A...1P}, and the right panels show member YSOs plotted against integrated CO data from \citet{2001ApJ...547..792D}. The CO data is
integrated through the velocity range of the optimal component identified using the procedure described in Sec. \ref{gas counterpart} 
 for Type 1, 2, 3, and 4 and
integrated over all the velocity channels for Type 5 and 6. The grid-like artifacts in
the CO map for Type 6 are caused by the incomplete sampling of the legacy observations.}
\label{matching example}
\end{center}
\end{figure*}

\subsection{Classification criteria} \label{criteria} 
The relationship between the spatial distribution of the YSOs and that of the gas can be diverse. We classify the YSO associations into 6 categories: 

\noindent (1) if the YSO association has a well-defined
counterpart seen in the CO map or the Planck dust map, it is classified as
Type 1 - direct YSO association; \\
\noindent (2) if the YSO association is located outside of a giant molecular cloud seen from the CO map or the Planck dust map, it is classified as
Type 2 - a close YSO association; \\
\noindent (3) if the YSOs in the association appear to stay inside a bubble-like structure in the CO map or the Planck dust map, it is classified as Type 3 - bubble YSO association; \\
\noindent (4) for large associations, the relationship between the YSO and gas is hard to quantify, as both the gas and the YSOs have complex, patchy distributions. This kind of association is considered a Type 4 - complex association.\\   
\noindent (5) if some diffuse gas appears in the vicinity of a YSO association, it's classified as Type 5 -  diffuse association; \\
\noindent (6) if no gas is detected around a YSO association, it's classified as Type 6 -  clean association; \\

We note that several
associations are hard to classify due to sparsely sampled data from legacy observations, and they are designated as unclassified (marked as Type 7 in the tables) and discarded in the later analysis. For Type 2 close and Type 3 bubble associations, we consider them as adjacent associations due to their relation with gas. A summary of the classification criteria can be found in Table. \ref{tablea}, and examples for each association type can be found in Fig. \ref{matching example}.

\begin{small}
\begin{table*} 
\caption{Classification criteria: criteria and information of different types of YSO associations \label{tablea}
}
\begin{tabular}{cccccc}

\hline
Type No.& Type & Criteria                                                                                                  & Association Number & Fraction (overall; per\,cent) & Size \\ \hline
1& Direct association              & Well associated with CO or dust cloud       &83            & 55.3  & cloud scale \\ 
2& Close association            &  Partially associated with CO or dust cloud         &    7    & 4.7 & cloud scale\\ 
3& Bubble association              &Associated with a bubble-like structure            &12      &8  & cloud scale\\ 
4& Complex association              & Associations with complex gas distributions                      &           19                & 12.7   & super cloud scale \\ 
5& Diffuse association     &Associated with some diffuse gas   & 16   & 10.7    & super cloud scale   \\ 
6& Clean association             &No counterpart in neither CO nor
                              dust map                &7           & 4.7  & super cloud scale\\    
   
7& Unclassified association          &Difficult to classify                                     &6   &  4   &  -   \\ \hline

\end{tabular}
\end{table*}
\end{small}

\section{Results} 
\subsection{Diversity}
The first discovery is that the YSO associations have a wide relation with nearby gas, as shown in Fig. \ref{matching example}. We classify these associations based on their relations with the gas.

Among all our associations, 80.7 \,per\,cent structures are related with their surrounding gas, which contains
the Type 1 direct associations\ (55.3\,per\,cent), Type 2 close associations\ (4.7\,per\,cent), Type 3 bubble
associations\ (8\,per\,cent) and Type 4 complex associations\ (12.7\,per\,cent). In these complex associations, part of them can be associated with gas while part of them are gas-poor. They are usually the maximum structure in our YSO associations. Nevertheless, we divide their substructures into different categories whenever possible.

Associations less related to gas take up 15.4\,per\,cent of all the
associations. These include structures where only a
diffuse gas component is detected (Type 5 diffuse associations: 10.7\,per\,cent) and structures
where no gas or dust counterpart exists (Type 6 clean association: 4.7\,per\,cent).

\subsection{Evolution}
\begin{figure*}
    \begin{center}
    \includegraphics[scale=0.25]{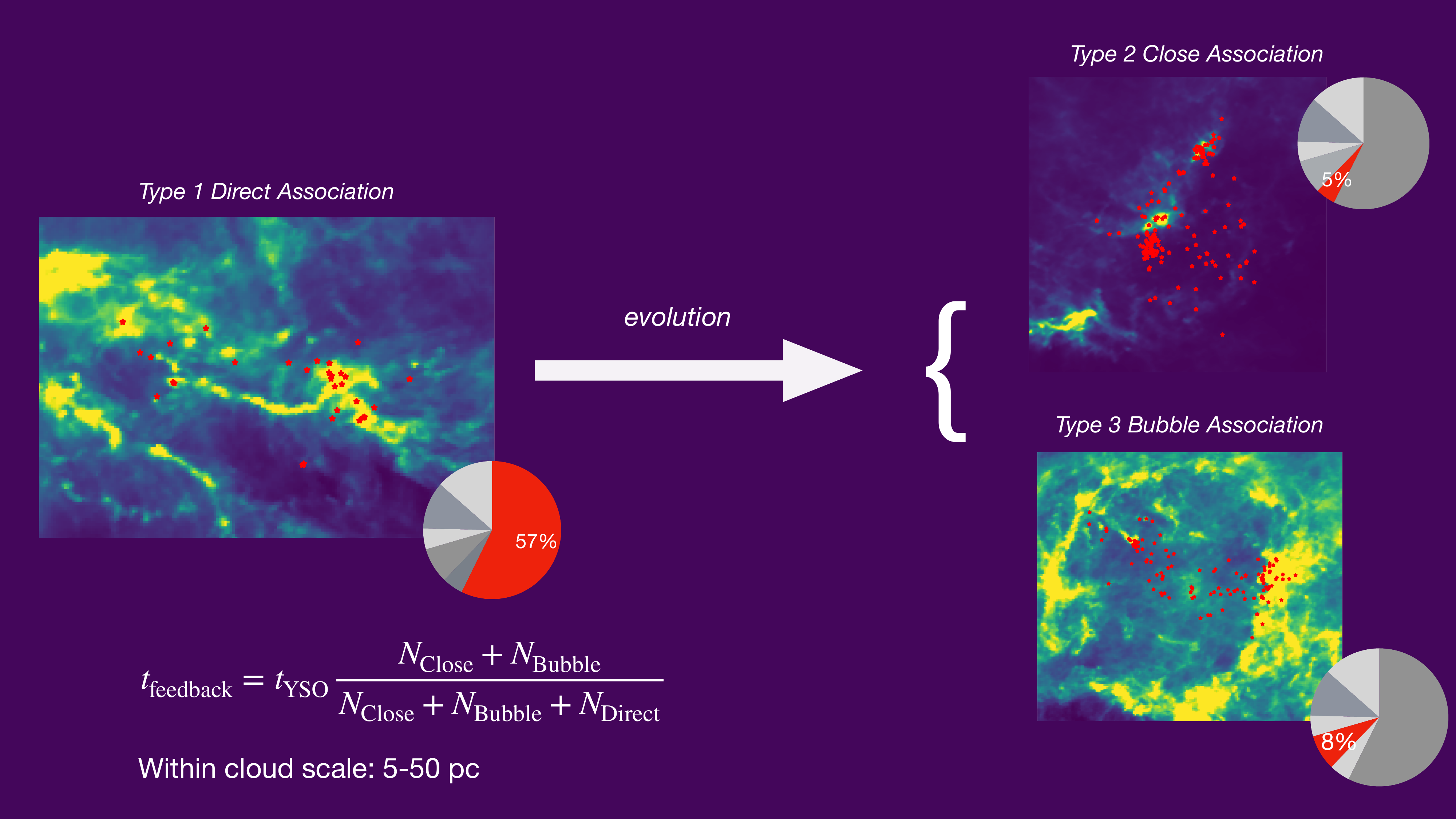}  
    \caption{{\bf YSO association classification and evolution 1 for cloud-scale associations.} Three inset images show the examples of Type 1 direct, Type 2 close, and Type 3 bubble associations. Member YSOs are plotted as red dots on the Planck dust map. The red part in each pie chart shows the fraction of this type in the whole sample. The arrow shows the proposed evolution sequence of the YSO associations.}
    \label{sequence1}
    \end{center}
    \end{figure*}

\begin{figure*}
    \begin{center}
    \includegraphics[scale=0.25]{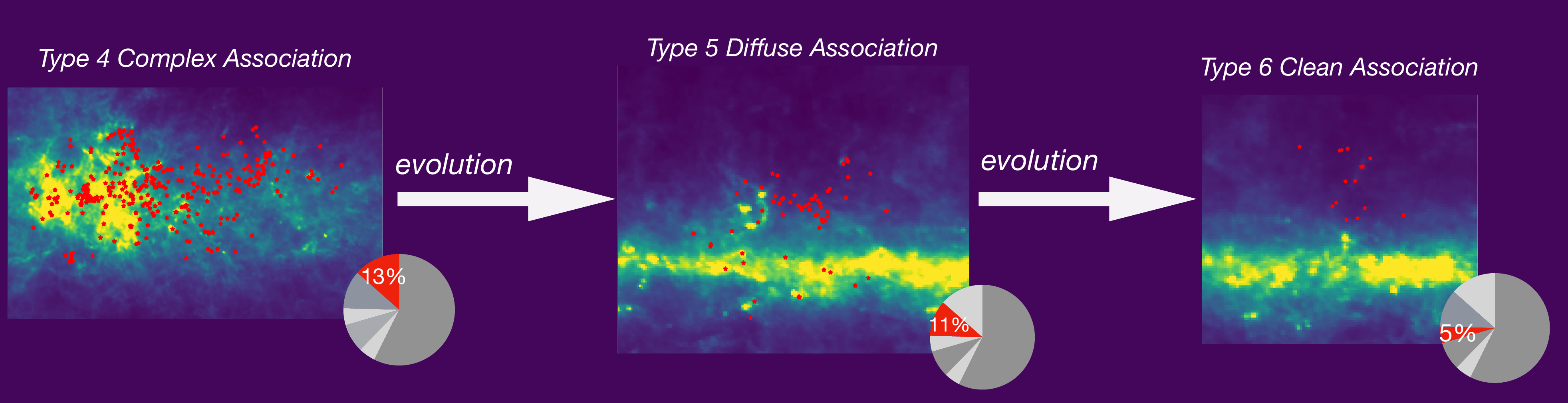}  
    \caption{{\bf YSO association classification and evolution 2 for some $r\approx 100\;\rm pc$ associations.} Three inset images show examples of Type 4 complex, Type 5 diffuse, and Type 6 clean associations. Member YSOs are plotted as red dots on the Planck dust map. The red part in each pie chart shows the fraction of this type in the whole sample. The arrows show the proposed evolution sequence of the YSO associations.}
    \label{sequence2}
    \end{center}
    \end{figure*}

\begin{figure}
    \begin{center}
    \includegraphics[scale=0.5]{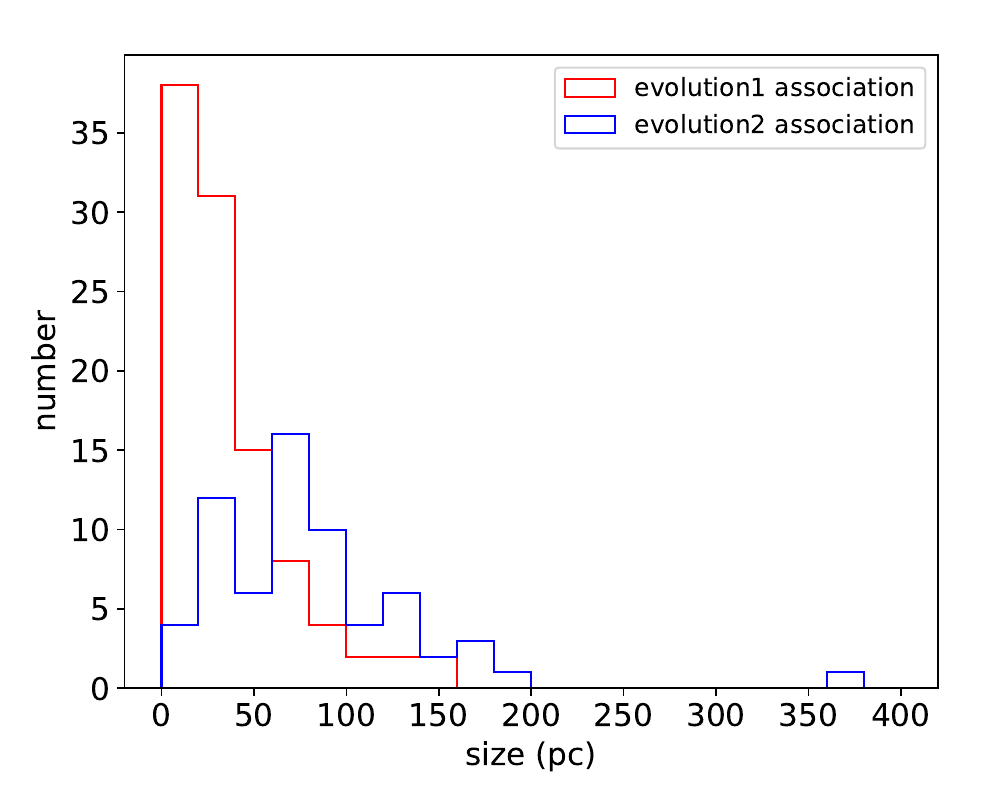}  
    \caption{{\bf Size distribution of YSO associations for evolution sequence 1 and 2.} The red histogram shows the size distribution for the associations used in evolution 1: Type 1 direct associations, Type 2 close associations, and Type 3 bubble associations. The blue histogram shows the size distribution for the associations in evolution 2: Type 4 complex associations, Type 5 diffuse associations, and Type 6 clean associations.}
    \label{distribution}
    \end{center}
    \end{figure}

Among the 6 types of YSO associations, based on the size of the structures, we present two separate evolution sequences for cloud-scale and supercloud-scale YSO associations in Fig. \ref{sequence1} and Fig. \ref{sequence2}. The size distributions for associations in these two evolution sequences are shown in Fig.\ref{distribution}. The mean size for evolution sequence 1 is 35.72 pc, and for evolution sequence 2 is 81.77 pc.

For the small, cloud-sized structures, the evolution sequence goes as follows (Fig. \ref{sequence1}, evolution 1): at the beginning, all the YSOs tend to be associated with molecular gas, making them the Type 1 direct association. As the young massive stars evolve in the association, the nearby gas is partially destructed or expelled by the stellar feedback, leading to Type 2 close associations and Type 3 bubble associations.

For some supercloud-size associations, we propose the following evolution (Fig. \ref{sequence2}, evolution 2): it starts with Type 4 complex associations, where the gas is still abundant, and these structures can evolve into Type 5 diffuse associations, and finally into type 6 clean associations. This distinction is still suggestive, and further studies are needed.

\subsection{Locations in the velocity-size ($\sigma_{v}-r$)
diagram} 
Turbulence is one of the most important controlling factors of molecular cloud evolution. A convenient way to study turbulence is to plot the velocity dispersions of different structures against their sizes. This plot was made by \citet[][]{1981MNRAS.194..809L} and has been called the "Larson relation" ever since. 
In \cite{2022MNRAS.513..638Z}, we derived the velocity dispersion-size relation towards all our YSO
associations, where we found a slope of 0.67 which is steeper than the 0.38 found in early studies
\citep[e.g.][]{1981MNRAS.194..809L}. 

\begin{figure*}
\begin{center}
\includegraphics[scale=0.9]{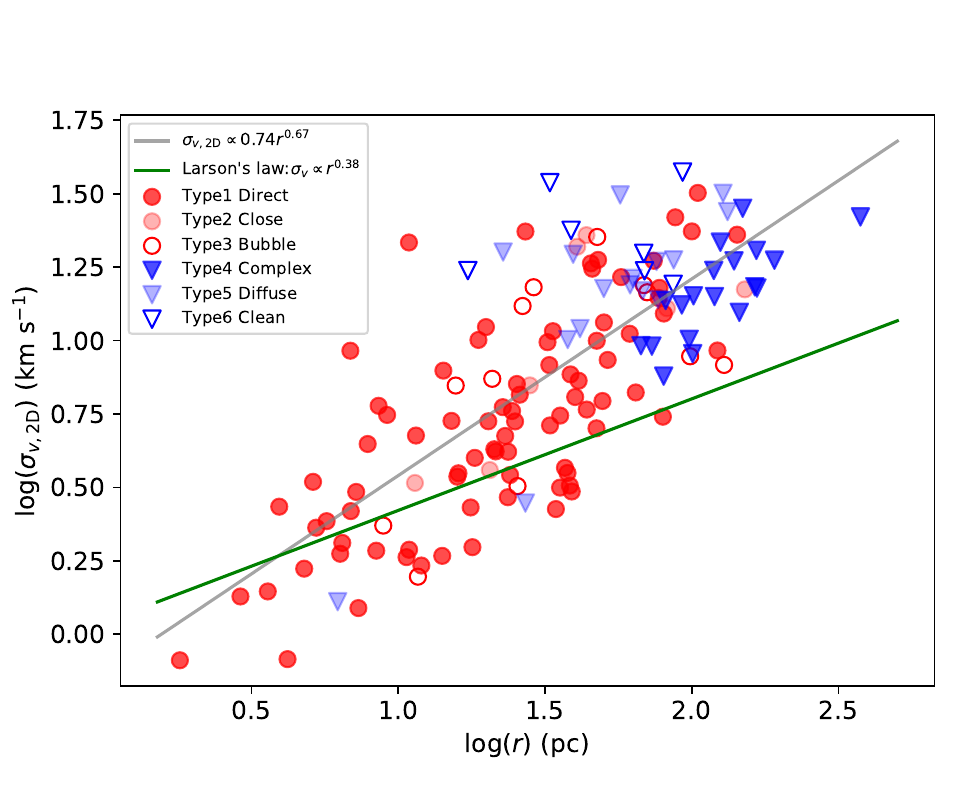} 
\caption{{\bf Locations of YSO associations of different types on $\sigma_{v}\
-\ r$ diagram.} Solid red circle, translucent red circle, and hollow red circle refer to Type 1 direct associations, Type 2 close associations, and Type 3 bubble associations respectively. Solid blue triangles, translucent blue triangles, and hollow triangles represent the Type 4 complex associations, Type 5 diffuse associations, and Type 6 clean associations respectively. The grey
line shows the velocity dispersion - size relation with fitting error added: $\sigma_{v, {\rm 2D}} = 10^{-0.13\pm 0.08}r^{0.67\pm 0.05}$ \citep{2022MNRAS.513..638Z}. The green line is the Larson's relation from \citet{1981MNRAS.194..809L}: $\sigma_{v} \propto r^{0.38}$.}
\label{larson}
\end{center}
\end{figure*}

In Fig. \ref{larson}, we plot the different types of associations on the velocity dispersion ($\sigma_v$) -
size ($r$) plane. 
The size and velocity dispersion information toward each association is present in Tab. \ref{result}. The sizes have been estimated in our last work \citep{2022MNRAS.513..638Z}. It refers to the FWHM (full width of half maximum) of the spatial distribution for the YSO association. $\sigma_v$ is the standard deviation of the 2D velocities for a certain YSO association using proper motion in $l$ and $b$ directions. 

In Fig. \ref{larson}, different types of YSO associations occupy
different parts of the diagram. At smaller scale (r $\lesssim$
50 pc), the majority of the YSO associations are either directly
associated with the gas (Type 1 direct association), or partially associated with the gas (Type 2 close \& 3 bubble association), and
they follow the relation of $\sigma_{v} \sim r^{0.68}$ as found by
\citet{2022MNRAS.513..638Z}, which is steeper than the green line found in \citet{1981MNRAS.194..809L}. On the large side, the velocity dispersions of the associations reach the range of $10 - 30$ $\rm km\;s^{-1}$. This size range is dominated by Type 4 complex, Type 5 diffuse, and Type 6 clean associations. Among these included in the evolution II sequence, there tends to be a velocity dispersion increase from Type 4 complex, Type 5 diffuse to Type 6 clean associations.

%Following the evolution sequence in Fig. \ref{sequence2}, from Type 4 %and Type 6 to Type 5 clean associations, there is an increase in the %velocity dispersion which can be explained by the energy injection %during the massive stellar feedback(\textcolor{red}{reference}).

%From Fig. \ref{larson}, gas-poor associations (Type 4 and 5) tend to %have a larger size than gas-related associations (Type 1, 2, and 3). In %our evolution picture, there is a gas removal process for the YSO %associations. The decrease in gas leads to the reduction of the potential %well in the association, causing the association expansion. The %expansion explains the larger size of gas-poor associations. This has %been found in previous studies and simulations %\citep{2007MNRAS.380.1589B, 2013A&A...559A..38P}.

\subsection{Locations in the Galaxy and on the sky}
\begin{figure*}
\begin{center}
\includegraphics[scale=0.4]{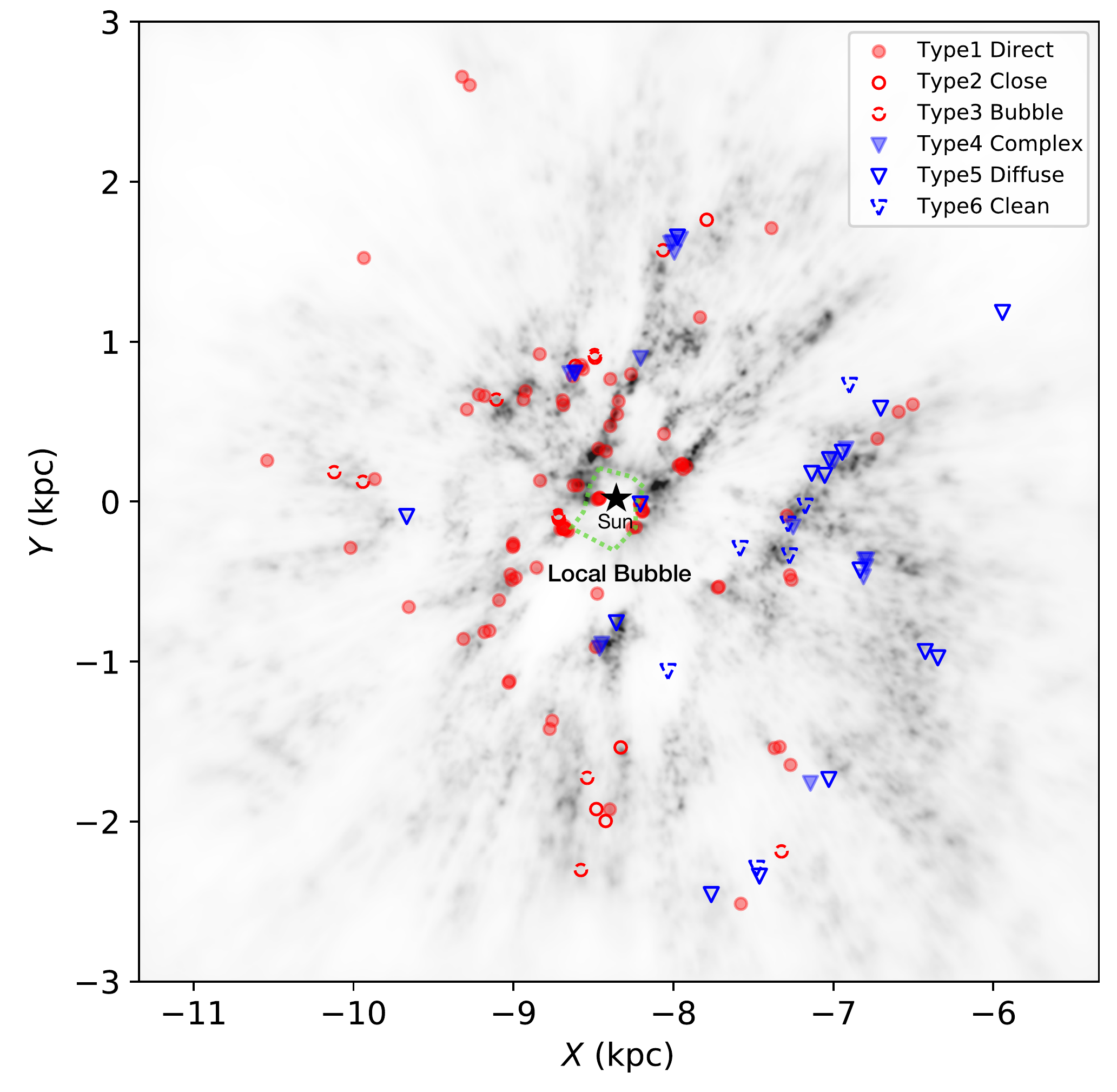} 
\caption{{\bf Locations of YSO associations of different types in the Galactic $XY$ plane}. The background is
the dust map from \citet{2019A&A...625A.135L}. Red circles and
blue triangles represent associations involved in evolution sequences 1 and 2, and different styles refer to different association types. The Galactic Center is located at (0,0) and the Sun
is located at ($-$8.34, 0) (black star). The green dashed lines mark several cavities in the solar neighborhood.}
\label{solar}
\end{center}
\end{figure*}

To study the distributions of different types of YSO associations in our Galaxy, we
use a three-dimensional dust map from \cite{2019A&A...625A.135L}. In Fig. \ref{solar}, we plot the dust distribution in the Galactic $X$-$Y$ plane, upon which the locations of the different types of YSO associations are overlaid. We notice that there is a group of Type 4 complex, Type 5 diffuse, and Type 6 clean associations distributed on the right of Fig. \ref{solar}, and most Type 1 direct, Type 2 close, and Type 3 Bubble associations are located on the two filaments near the Sun. We mark these filaments in Fig. \ref{clean} and make a discussion in Sec. \ref{filaments}

\begin{figure*}
\begin{center}
\includegraphics[scale=0.275]{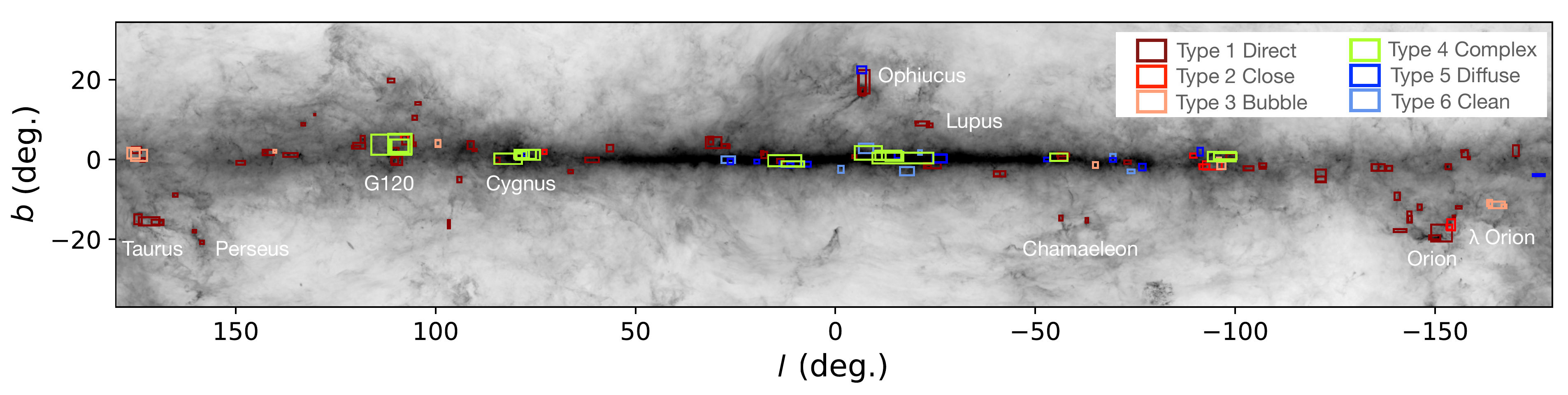}  
\caption{{\bf Locations of different types of YSO associations on the sky.} The background is the dust map from \citet{2014A&A...571A...1P}. Dark red, red, orange, lemon yellow, blue, and light blue squares refer to Type 1 direct associations, Type 2 close associations, Type 3 bubble associations, Type 4 complex associations, Type 5 diffuse associations, and Type 6 clean associations respectively. The sizes of the squares represent the areas that the YSO associations cover in the sky plane. The locations of some famous molecular clouds are indicated.}
\label{planck}
\end{center}
\end{figure*}

In Fig. \ref{planck}, we  plot the locations of different types of YSO associations on 
the Planck dust map to show their location on the sky plane. At higher Galactic latitudes, our YSO associations match with some
well-known molecular clouds like  Perseus, Taurus, and Orion molecular clouds. We also recovered some small clouds at high latitudes. The majority of our YSO associations stay close to the Galactic mid-plane. 

% Being huge and complex associations, they can provide unique information for
% the giant star formation region. From the survey targeting at these
% associations on the Galactic plane, we can have a better view of the gas
% dynamics and star formation at large scale. 

\subsection{Gas mass and virial parameter}
In Sec. \ref{gas counterpart}, we find the CO counterpart for most YSO
associations, especially for gas-rich associations: Type 1 direct and Type 4 complex associations. This gas-rich property provides us with an opportunity to study the gas mass in the associations. As the widely used tracer for H$_2$ molecular gas, CO emission can
be converted into the column density of Hydrogen gas using the factor $X$(CO).
We adopt $X({\rm CO}) = 2\times10^{20}\ {\rm cm}^{-2} ({\rm K}\ {\rm km}\ {\rm
s}^{-1})^{-1}$ \citep{2013ARA&A..51..207B}. The gas mass is calculated using:
${M} = I_{\rm CO} \times X(\rm CO) \times m_{\rm H_2} \times 1.35 A$, where
$I_{\rm CO}$ is the CO emission, $m_{\rm H_2}$ is the molecular mass of Hydrogen
and A is the area defined in by $[l_{{\rm mean}, i} \pm 3{\rm std}{(l_i)}, b_{{\rm mean}, i}
\pm 3{\rm std}{(b_i)}]$, where $l_{{\rm mean}, i}$ and $b_{{\rm mean}, i}$ are the mean galactic latitude and longitude for association $i$, std$(l_i)$ and std$(b_i)$ are the standard deviations of galactic latitudes and longitudes for association $i$. The multiple factor 1.35 is due to the metallicity of the
molecular gas. To calculate the gas mass without background contamination, we
just use the emission of the chosen gas component (more details about CO components in Appendix. \ref{appendixa}). We integrate the CO emission
of the gas counterpart between $\overline{v_i}-{\rm std}(v_i)$ and
$\overline{v_i}+{\rm std}(v_i)$, where $\overline{v_i}$ is the centroid velocity for association $i$ and
${\rm std} (v_i)$ is the velocity dispersion of the
associated component for association $i$.   

\begin{figure}
\begin{center}
\includegraphics[scale=0.5]{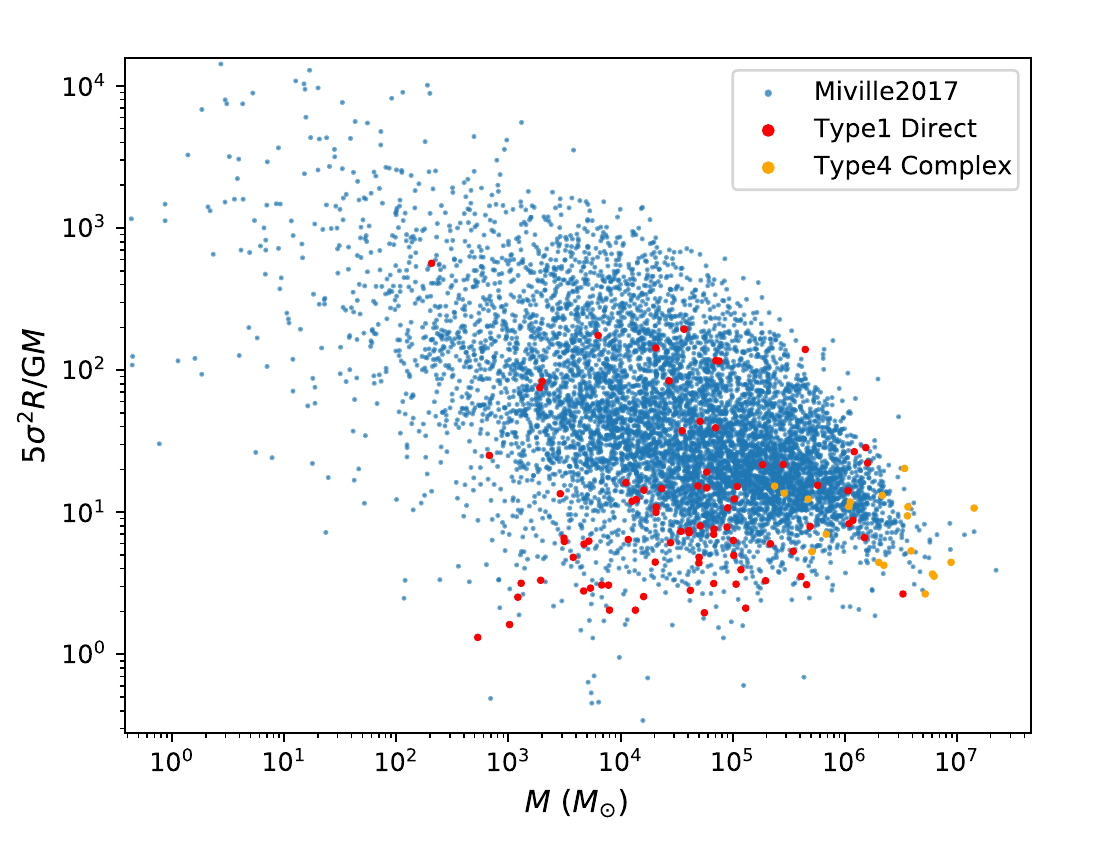}  
\caption{{\bf Viral parameter vs. gas mass for Type 1 direct and Type 4 complex YSO associations.} Blue dots are the clouds from \citet{2017ApJ...834...57M}. Red dots refer to the Type 1 direct associations and Orange dots refer to Type 4 complex associations.} 
\label{virial}
\end{center}
\end{figure}

After having derived gas mass for those YSO associations, we study the importance of gravity in the molecular gas by estimating the virial parameter $\alpha_{\rm vir} = 5\sigma^2R/{{\rm G}M}$ \citep{1992ApJ...395..140B}, where $\sigma$ is the 1D velocity dispersion ($\sigma = \sigma_{\rm 2D}/\sqrt{2}$, using our $\sigma_{\rm 2D}$ derived from YSO proper motion in $l$ and $b$ directions), $R$ is the size of the YSO association, $M$ is the cloud mass derived from CO emission and $G$ is the gravitational constant. The virial parameter describes the ratio between kinetic energy and gravitational energy of a molecular cloud. A cloud with $\alpha\ \le\ 1$ is gravity-dominated and is likely to collapse. We plot the virial parameter versus the gas mass in Fig. \ref{virial}, where nearly all presented types of YSO associations have virial parameters larger than 1, showing subcritical properties. This is consistent with previous studies that molecular clouds tend to be gravitationally unbound \citep{2009ApJ...699.1092H, 2011MNRAS.413.2935D, 2017ApJ...834...57M}. All the virial parameters and mass values can be found in Tab. \ref{result}. In Fig. \ref{virial}, we compare the virial parameter of our gas-rich sample with that of 8107 molecular clouds from \citet{2017ApJ...834...57M}. \citet{2017ApJ...834...57M} found a modest correlation between the virial parameter and the cloud mass in their sample: $\alpha_{\rm virial} \propto M^{-0.53\pm0.3}$. Although this correlation can not be seen in our Type 1 and Type 4 associations, our sample covers a wide range in the virial paramater-mass diagram.

\section{Discussions}

\subsection{Timescale of gas displacement by stellar feedback}
Although YSOs are very young objects born from molecular clouds, and one would
naturally expect them to be spatially associated with gas, we observe some associations
that are associated but not fully associated with gas. These include the Type 2 close and Type 3 bubble associations. In those cases, the member YSOs are located at the edge of clouds
or inside bubbles. We propose that the displacement and decreasing gas fraction are likely to be the results for Type 1 direct associations after gas removal by the massive stellar feedback. The fraction of feedback-affected
YSO associations (Type 2 close \& Type 3 bubble associations) can be used to infer the feedback timescale. We set a size range of 5$-$50 pc for the associations used in the calculation. 
Towards the association numbers with sizes ranging between 5 - 50 pc, there are 62 Type 1 associations, 5 Type 2 associations, and 8 Type 3 associations in the sample, respectively. Considering the uncertainty of the YSO age, we adopt an upper limit lifetime of 10 Myr for YSOs. 
We use the ratio between
different types among gas-related associations to estimate the time that takes for stellar
feedback to displace gas:

\begin{small}
\begin{equation}
t_{\rm feedback} = t_{\rm YSO}\times \frac{ n_{\rm Type 2} + n_{\rm Type 3}}{n_{\rm Type 1} + n_{\rm Type 2} + n_{\rm Type 3}} \approx 10\; {\rm Myr}\; \times\, 0.17 \approx 1.7 \; \rm Myr \;.
\end{equation} 
\end{small}

This upper limit of the feedback time is comparable to a short disruption time of about 1.5 Myr
estimated by \cite{2019Natur.569..519K} towards a face-on, star-forming disc
galaxy NGC300. Compared to the typical age of several million years of YSO
\citep{2015ApJS..220...11D} and a 30-Myr lifetime of molecular clouds
\citep{1977ApJ...217..464B, 1991ASIC..342...35E, 1994LNP...439...13L,
2009ApJS..184....1K}, the feedback time, which is around 17\% of the upper limit of the YSO time, is quite short. A timeline is summarized
in Fig.\ref{sequence1}. Considering the location of our sample, this estimation only represents the feedback time in the solar neighborhood.

\subsection{Gas-free YSO associations}
Our analyses also reveal several Type 6 clean associations that contain little or no gas.
Their location can be seen in Fig. \ref{solar} and
Fig. \ref{clean}.
This lack of gas is examined and confirmed again using the 3D dust map from
\cite{2019MNRAS.483.4277C}, which allows us to compare the YSO association with
dust distribution in 3D. 

Type 6 clean associations represent a class of YSO associations where
the gas of the ambient cloud is expelled. From a map of the locations of these associations in
our Galaxy (Fig.\ref{clean}), we find that many gas-poor
associations reside in large, superbubble-like cavities, consistent with their gas-poor nature

These gas-free structures occupy a higher part of the velocity
dispersion size when compared to the  Type 4 complex and Type 5 diffuse associations at similar sizes. This
higher velocity dispersion of the gas-free indicates energy injection during the evolution, which can be caused by gas removal and stellar feedback.

Simulations studying the effect of residual-gas expulsion in \citet{2007MNRAS.380.1589B} showed radially anisotropic velocity dispersions increase for star clusters with gas removed within the initial crossing time.
In \citet{2020ApJ...900L...4P}, they have found the increasing trend of the 2D velocity dispersion in some young star clusters under the gas expulsion. 
This has been mentioned when \citet{2006MNRAS.373..752G, 2016A&A...587A..53K} studied the gas expulsion in the young star clusters. 
Also, the increased velocity dispersion is found
in simulations when there is supernovae and stellar feedback added \cite[e.g.][]{2022MNRAS.513.2088B}.

\subsection{Variations across Galactic-scale filaments}\label{filaments}

\begin{figure}
\begin{center}
\includegraphics[scale=0.25]{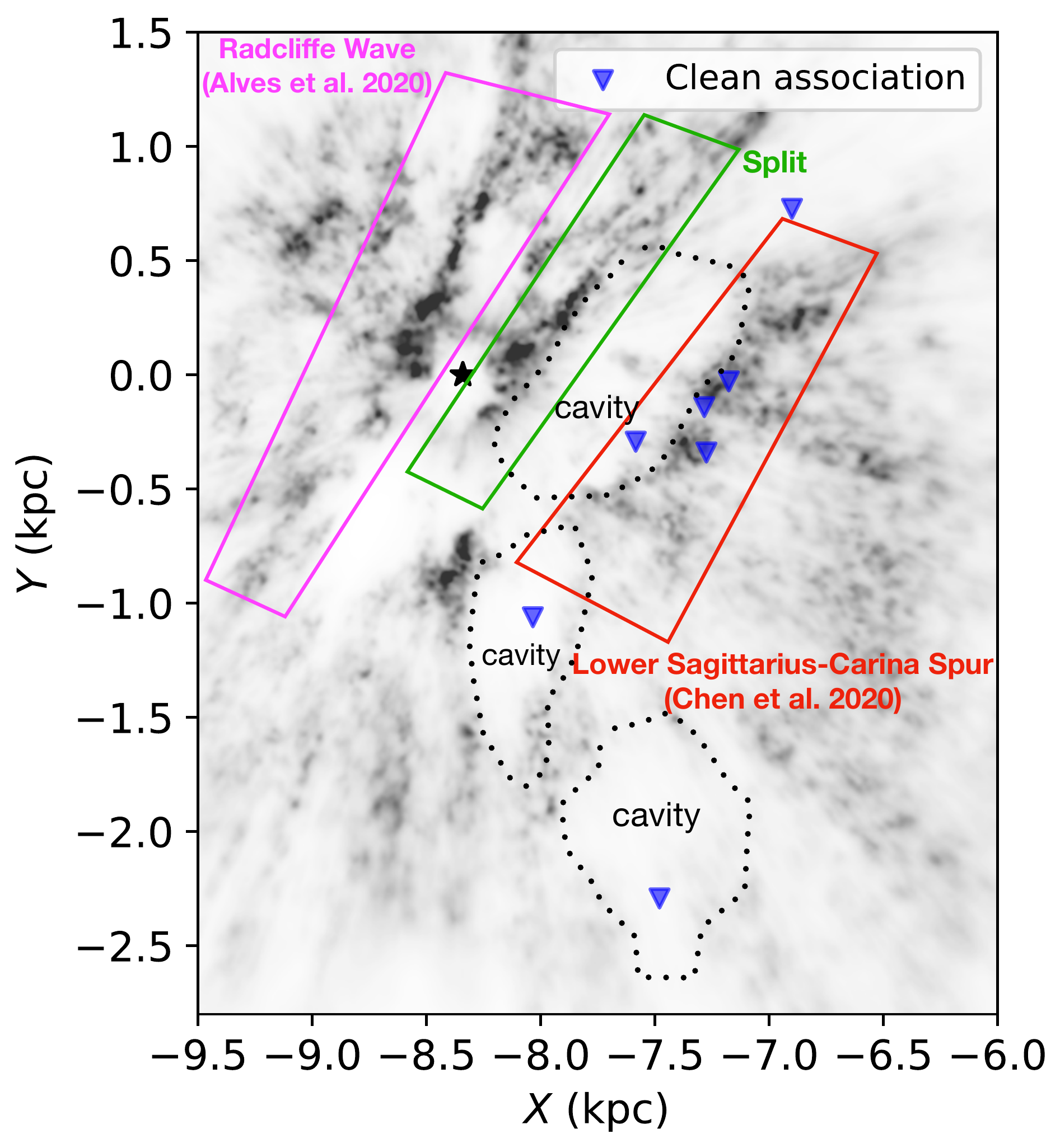} 
\caption{{\bf Locations of clean associations on $X$ - $Y$ plane.} The background is a dust map from \citet{2019A&A...625A.135L}. The Sun is located at ($-$8.34\,kpc, 0) marked by the black star. Blue triangles refer to the clean associations (Type 6). The pink, green, and red squares mark the Radcliffe Wave filament, Split, and Lower Sagittarius-Carina Spur, respectively. The black dashed lines mark several cavities in the solar neighborhood.}
\label{clean}
\end{center}
\end{figure}

Gas in the Milky Way is organized in kpc-sized filaments. Early studies like \cite{2013A&A...559A..34L}, \cite{2014ApJ...797...53G} and \cite{2014A&A...568A..73R} have revealed these giant coherent molecular structures in observations. \cite{2020Natur.578..237A} and \cite{2022arXiv220503218L} have studied a 2.7-kpc dense filament structure called Radcliffe Wave in solar neighborhood (pink square in Fig. \ref{clean}), which contains most of the nearby star-forming regions.

In Fig. \ref{clean}, we plot the distribution of dust in the Galactic disk plane, where we mark three Galactic-scale filament structures, including the Radcliffe Wave, a filament studied in \cite{2020MNRAS.493..351C} (Lower Sagittarius-Carina Spur) and a smaller filament between them, called the "Split" in \cite{2019A&A...625A.135L}. Combined with Fig. \ref{solar},
we found that the gas-richness and association types of different Galactic-scale filaments vary significantly. For example, the Radcliffe Wave contains mostly gas-rich associations, yet the Lower Sagittarius-Carina Spur filament \citep{2020MNRAS.493..351C,2021A&A...651L..10K}, which is a kpc-size gas filament located at the Sagittarius arm, containing a significantly higher fraction diffuse, or clean YSO associations, indicating a lot of stellar feedback. More studies about star formation and filament structures and properties are needed to help us better understand the difference between these filaments.

\section{Conclusions}
Combining the YSO associations, CO data and Planck dust map, we study the surrounding gas for a sample of $150$ YSO associations and classify these associations based on their relation with the gas. Our  findings include:

%  We find that these YSO associations exhibit a
%  wide range of gas contents. 
%  Our sample includes YSO associations which have
%  well-defined molecular cloud counterparts and both little gas. Based on the gas
%  contents, we divide our YSO associations into 3 different types (Type 1:
%  gas-rich, Type 2: gas-poor, Type 3: huge YSO associations), and a total of six
%  subtypes (Type 1.1: direct, Type 1.2: close, Type 1.3: bubble, Type 2.1:
%  diffuse, Type 2.2: clean, Type 3: huge associations). We also found large
%  coherent associations classified as huge complexes. 

\begin{table*}
    \centering
    \caption{Lists of information of the 150 YSO associations. The table contains location, association types based on relation with nearby gas, 2D velocity dispersion, size, mass, virial parameter and YSO number. The full table can be downloaded online through ...}
    \label{result}
    \begin{tabular}{cccccccccc}
    \hline
    id  & Galactic ID      & $l$ ($^\circ$)     & $b$ ($^\circ$)     & type & $\sigma_{v,{\rm 2D}}$ (km s$^{-1}$) & size (pc)      & mass ($M_{\odot}$)       & virial parameter  & YSO number \\\hline
    0   & G171.700-015.500 & 171.67 & -15.46 & 1    & 1.94       & 10.90  & 7803.19     & 3.07   & 62     \\
    1   & G169.600-015.700 & 169.56 & -15.70 & 1    & 1.88       & 6.35   & 4690.81     & 2.79   & 33     \\
    2   & G168.700-015.900 & 168.68 & -15.88 & 1    & 1.34       & 2.90   & 1213.79     & 2.52   & 24     \\
    3   & G174.500-015.000 & 174.47 & -15.00 & 1    & 2.05       & 6.45   & 5393.88     & 2.92   & 45     \\
    4   & G338.205+009.000 & 338.21 & 8.99   & 1    & 1.92       & 8.42   & 3784.53     & 4.81   & 47     \\
    5   & G336.422+008.600 & 336.42 & 8.56   & 1    & 2.72       & 3.94   & 675.97      & 25.11  & 21     \\
    6   & G339.665+009.300 & 339.67 & 9.35   & 1    & 0.82       & 1.81   & 532.72      & 1.31   & 24     \\
    7   & G352.880+019.500 & 352.88 & 19.48  & 1    & 1.85       & 14.13  & 4721.85     & 5.97   & 255    \\
    8   & G353.255+018.900 & 353.26 & 18.91  & 1    & 1.71       & 11.99  & 3129.27     & 6.57   & 166    \\
    9   & G353.199+017.000 & 353.20 & 17.01  & 1    & 1.67       & 4.79   & 0.00        & inf    & 107    \\
    10  & G353.383+022.500 & 353.38 & 22.47  & 5    & 1.29       & 6.23   & 0.00        &  -      & 68     \\
    11  & G297.100-015.200 & 297.06 & -15.25 & 1    & 1.40       & 3.60   & 1298.95     & 3.16   & 59     \\
    12  & G303.600-014.700 & 303.58 & -14.67 & 1    & 0.82       & 4.20   & 1024.04     & 1.62   & 24     \\
    13  & G158.400-020.700 & 158.45 & -20.74 & 1    & 2.43       & 5.71   & 3151.54     & 6.22   & 49     \\
    14  & G160.300-017.900 & 160.34 & -17.94 & 1    & 3.30       & 5.13   & 5217.63     & 6.26   & 33     \\
    15  & G104.424+014.000 & 104.42 & 14.02  & 1    & 1.23       & 7.32   & 1939.91     & 3.32   & 18     \\
    ...  & ... & ... & ...  & ...  & ... & ... & ... & ... & ...\\
    \hline

\end{tabular}
\end{table*}

\begin{enumerate}
    \item {\bf Diversity}. The YSO association sample sources have diverse relations with the surrounding gas. Based on their relations with gas, they are classified into six types: (1) Type 1 direct associations with clear gas counterpart; (2) Type 2 close associations partially associated with gas; (3) Type 3 bubble associations partially associated with bubble-like gas structures; (4) Type 4 complex associations with complex gas counterparts. Type 4 tends to be the parent structures for some smaller structures; (5) Type 5 diffuse associations with very diffuse gas counterparts; (6) Type 6 clean associations with no gas counterpart.

   \item {\bf Evolution of cloud-sized ($l \approx 30\;\rm  pc$) objects.} The different gas-richness results from evolution. From Type 1 direct associations to Type 2 close and Type 3 bubble associations, the gas displacement is likely caused by stellar feedback of the massive stars. 
   \item  {\bf Evolution of $l \gtrsim 100\;\rm pc$ objects.}  We found an evolution sequence from Type 4 complex, Type 5 diffuse to Type 6 clean associations. This sequence also shows the gas decrease in the associations due to the stellar feedback.

   \item {\bf Different types occupy different Locations in the velocity-size plane.} The gas-rich Type 1 direct, Type 2 close, and Type 3 bubble associations follow a velocity-size relation which is consistent with our previous results. 
   
   For associations in evolution 2, going from the Type 4 complex, Type 5 diffuse to Type 6 clean YSO associations, there is a continuous increase in the velocity dispersion, indicating the energy injection during stellar feedback. 
   
   % The additional velocity dispersion can also be generated during the gas removal.
   
   \item {\bf Short feedback time and rapid gas removal.} From the number of feedback-related types of YSO associations within 5-50 pc, we derive an upper limit timescale of around 1.7 Myr ($t_{\rm feedback} < 0.17\;  t_{\rm YSO}$) as the time it takes for stellar feedback to remove gas from a cloud. 
   
   \item  {\bf Gas-free YSO associations.} We also discovered a population of gas-free YSO associations. They are located either inside kpc-sized superbubble-like cavities or on a Galactic-scale diffuse filament. These gas-free YSO associations are at a very late stage of evolution where gas has been removed. These structures might later evolve into other forms of the concentrations of young stars, such as the Gaia stellar strings \citep{2020AJ....160..279K}.
   
    \item {\bf Different association types on different filaments.} We found the YSO association types differ across different filament structures, indicating different stages of the filaments. More detailed studies are needed. 

   \item {\bf Gas-rich structures are gravitationally unbound.} By computing the virial parameter for Type 1 direct and Type 4 complex associations where most gas has not yet been removed, we find that nearly all these YSO associations have $\alpha_{\rm vir} > 1$, indicating they are gravitationally unbound, consistent with previous studies.    
   
   Our sample sets a solid foundation to explore the relation
   between interstellar medium evolution, star formation, and Galaxy structure, and this potential will be exploited in our future papers. 
   
\end{enumerate}

\section*{Acknowledgements}
We thank our referee for a careful reading of the paper and the constructive comments. 
This work is partially supported by the Post-graduate Research and Innovation Project of Yunnan University (No. 2019236) and the China Scholarship Council (CSC).
GXL acknowledges support from NSFC grant No. 12273032 and
12033005. BQC is supported by the National Key R\&D Program of China No. 2019YFA0405500, National Natural Science Foundation of China 12173034, 11803029 and 11833006, and the science research
grants from the China Manned Space Project with NO.\, CMS-CSST-2021-A09, CMS-CSST-2021-A08 and CMS-CSST-2021-B03.

This work presents results from the European Space Agency (ESA) space mission
Gaia. Gaia data are being processed by the Gaia Data Processing and Analysis
Consortium (DPAC). Funding for the DPAC is provided by national institutions, in
particular, the institutions participating in the Gaia Multilateral Agreement
(MLA). The Gaia mission website is https://www.cosmos.esa.int/gaia. The Gaia
archive website is https://archives.esac.esa.int/gaia. This research has used
ASTRODENDRO, a PYTHON package to compute dendrograms of Astronomical data
(http://www.dendrograms.org/).

\section*{DATA AVAILABILITY}
The paper makes use of published data from  \citet{2016MNRAS.458.3479M, 2022MNRAS.513..638Z}, Gaia DR2  \citep{2016A&A...595A...1G}. Our Table \ref{result} containing the location and classification type information will be made available online upon publication.

% \section*{Author contributions}

\bibliographystyle{mnras}
\bibliography{class_new.bib}

\begin{thebibliography}{}
\makeatletter
\relax
\def\mn@urlcharsother{\let\do\@makeother \do\$\do\&\do\#\do\^\do\_\do\%\do\~}
\def\mn@doi{\begingroup\mn@urlcharsother \@ifnextchar [ {\mn@doi@} {\mn@doi@[]}}
\def\mn@doi@[#1]#2{\def\@tempa{#1}\ifx\@tempa\@empty \href {http://dx.doi.org/#2} {doi:#2}\else \href {http://dx.doi.org/#2} {#1}\fi \endgroup}
\def\mn@eprint#1#2{\mn@eprint@#1:#2::\@nil}
\def\mn@eprint@arXiv#1{\href {http://arxiv.org/abs/#1} {{\tt arXiv:#1}}}
\def\mn@eprint@dblp#1{\href {http://dblp.uni-trier.de/rec/bibtex/#1.xml} {dblp:#1}}
\def\mn@eprint@#1:#2:#3:#4\@nil{\def\@tempa {#1}\def\@tempb {#2}\def\@tempc {#3}\ifx \@tempc \@empty \let \@tempc \@tempb \let \@tempb \@tempa \fi \ifx \@tempb \@empty \def\@tempb {arXiv}\fi \@ifundefined {mn@eprint@\@tempb}{\@tempb:\@tempc}{\expandafter \expandafter \csname mn@eprint@\@tempb\endcsname \expandafter{\@tempc}}}

\bibitem[\protect\citeauthoryear{{Abt}}{{Abt}}{1983}]{1983ARA&A..21..343A}
{Abt} H.~A.,  1983, \mn@doi [\araa] {10.1146/annurev.aa.21.090183.002015}, \href {https://ui.adsabs.harvard.edu/abs/1983ARA&A..21..343A} {21, 343}

\bibitem[\protect\citeauthoryear{{Adams}, {Lada}  \& {Shu}}{{Adams} et~al.}{1987}]{1987ApJ...312..788A}
{Adams} F.~C.,  {Lada} C.~J.,   {Shu} F.~H.,  1987, \mn@doi [\apj] {10.1086/164924}, \href {https://ui.adsabs.harvard.edu/abs/1987ApJ...312..788A} {312, 788}

\bibitem[\protect\citeauthoryear{{Allen} et~al.,}{{Allen} et~al.}{2004}]{2004ApJS..154..363A}
{Allen} L.~E.,  et~al., 2004, \mn@doi [\apjs] {10.1086/422715}, \href {https://ui.adsabs.harvard.edu/abs/2004ApJS..154..363A} {154, 363}

\bibitem[\protect\citeauthoryear{{Alves} et~al.,}{{Alves} et~al.}{2020}]{2020Natur.578..237A}
{Alves} J.,  et~al., 2020, \mn@doi [\nat] {10.1038/s41586-019-1874-z}, \href {https://ui.adsabs.harvard.edu/abs/2020Natur.578..237A} {578, 237}

\bibitem[\protect\citeauthoryear{{Bash}, {Green}  \& {Peters}}{{Bash} et~al.}{1977}]{1977ApJ...217..464B}
{Bash} F.~N.,  {Green} E.,   {Peters} W.~L. I.,  1977, \mn@doi [\apj] {10.1086/155595}, \href {https://ui.adsabs.harvard.edu/abs/1977ApJ...217..464B} {217, 464}

\bibitem[\protect\citeauthoryear{{Baumgardt} \& {Kroupa}}{{Baumgardt} \& {Kroupa}}{2007}]{2007MNRAS.380.1589B}
{Baumgardt} H.,  {Kroupa} P.,  2007, \mn@doi [\mnras] {10.1111/j.1365-2966.2007.12209.x}, \href {https://ui.adsabs.harvard.edu/abs/2007MNRAS.380.1589B} {380, 1589}

\bibitem[\protect\citeauthoryear{{Bending}, {Dobbs}  \& {Bate}}{{Bending} et~al.}{2022}]{2022MNRAS.513.2088B}
{Bending} T. J.~R.,  {Dobbs} C.~L.,   {Bate} M.~R.,  2022, \mn@doi [\mnras] {10.1093/mnras/stac965}, \href {https://ui.adsabs.harvard.edu/abs/2022MNRAS.513.2088B} {513, 2088}

\bibitem[\protect\citeauthoryear{{Bertoldi} \& {McKee}}{{Bertoldi} \& {McKee}}{1992}]{1992ApJ...395..140B}
{Bertoldi} F.,  {McKee} C.~F.,  1992, \mn@doi [\apj] {10.1086/171638}, \href {https://ui.adsabs.harvard.edu/abs/1992ApJ...395..140B} {395, 140}

\bibitem[\protect\citeauthoryear{{Bolatto}, {Wolfire}  \& {Leroy}}{{Bolatto} et~al.}{2013}]{2013ARA&A..51..207B}
{Bolatto} A.~D.,  {Wolfire} M.,   {Leroy} A.~K.,  2013, \mn@doi [\araa] {10.1146/annurev-astro-082812-140944}, \href {https://ui.adsabs.harvard.edu/abs/2013ARA&A..51..207B} {51, 207}

\bibitem[\protect\citeauthoryear{{Chen} et~al.,}{{Chen} et~al.}{2019}]{2019MNRAS.483.4277C}
{Chen} B.~Q.,  et~al., 2019, \mn@doi [\mnras] {10.1093/mnras/sty3341}, \href {https://ui.adsabs.harvard.edu/abs/2019MNRAS.483.4277C} {483, 4277}

\bibitem[\protect\citeauthoryear{{Chen} et~al.,}{{Chen} et~al.}{2020}]{2020MNRAS.493..351C}
{Chen} B.~Q.,  et~al., 2020, \mn@doi [\mnras] {10.1093/mnras/staa235}, \href {https://ui.adsabs.harvard.edu/abs/2020MNRAS.493..351C} {493, 351}

\bibitem[\protect\citeauthoryear{{Chevance}, {Krumholz}, {McLeod}, {Ostriker}, {Rosolowsky}  \& {Sternberg}}{{Chevance} et~al.}{2022}]{2022arXiv220309570C}
{Chevance} M.,  {Krumholz} M.~R.,  {McLeod} A.~F.,  {Ostriker} E.~C.,  {Rosolowsky} E.~W.,   {Sternberg} A.,  2022, arXiv e-prints, \href {https://ui.adsabs.harvard.edu/abs/2022arXiv220309570C} {p. arXiv:2203.09570}

\bibitem[\protect\citeauthoryear{{Dame}, {Hartmann}  \& {Thaddeus}}{{Dame} et~al.}{2001}]{2001ApJ...547..792D}
{Dame} T.~M.,  {Hartmann} D.,   {Thaddeus} P.,  2001, \mn@doi [\apj] {10.1086/318388}, \href {https://ui.adsabs.harvard.edu/abs/2001ApJ...547..792D} {547, 792}

\bibitem[\protect\citeauthoryear{{Dobbs}, {Burkert}  \& {Pringle}}{{Dobbs} et~al.}{2011}]{2011MNRAS.413.2935D}
{Dobbs} C.~L.,  {Burkert} A.,   {Pringle} J.~E.,  2011, \mn@doi [\mnras] {10.1111/j.1365-2966.2011.18371.x}, \href {https://ui.adsabs.harvard.edu/abs/2011MNRAS.413.2935D} {413, 2935}

\bibitem[\protect\citeauthoryear{{Draine}}{{Draine}}{2011}]{2011piim.book.....D}
{Draine} B.~T.,  2011, {Physics of the Interstellar and Intergalactic Medium}

\bibitem[\protect\citeauthoryear{{Dunham} et~al.,}{{Dunham} et~al.}{2015}]{2015ApJS..220...11D}
{Dunham} M.~M.,  et~al., 2015, \mn@doi [\apjs] {10.1088/0067-0049/220/1/11}, \href {https://ui.adsabs.harvard.edu/abs/2015ApJS..220...11D} {220, 11}

\bibitem[\protect\citeauthoryear{{Elmegreen}}{{Elmegreen}}{1991}]{1991ASIC..342...35E}
{Elmegreen} B.~G.,  1991, in {Lada} C.~J.,  {Kylafis} N.~D.,  eds,  NATO Advanced Study Institute (ASI) Series C Vol. 342, The Physics of Star Formation and Early Stellar Evolution. p.~35

\bibitem[\protect\citeauthoryear{{Fall}, {Krumholz}  \& {Matzner}}{{Fall} et~al.}{2010}]{2010ApJ...710L.142F}
{Fall} S.~M.,  {Krumholz} M.~R.,   {Matzner} C.~D.,  2010, \mn@doi [\apjl] {10.1088/2041-8205/710/2/L142}, \href {https://ui.adsabs.harvard.edu/abs/2010ApJ...710L.142F} {710, L142}

\bibitem[\protect\citeauthoryear{{Feigelson} \& {Montmerle}}{{Feigelson} \& {Montmerle}}{1999}]{1999ARA&A..37..363F}
{Feigelson} E.~D.,  {Montmerle} T.,  1999, \mn@doi [\araa] {10.1146/annurev.astro.37.1.363}, \href {https://ui.adsabs.harvard.edu/abs/1999ARA&A..37..363F} {37, 363}

\bibitem[\protect\citeauthoryear{{Findeisen} \& {Hillenbrand}}{{Findeisen} \& {Hillenbrand}}{2010}]{2010AJ....139.1338F}
{Findeisen} K.,  {Hillenbrand} L.,  2010, \mn@doi [\aj] {10.1088/0004-6256/139/4/1338}, \href {https://ui.adsabs.harvard.edu/abs/2010AJ....139.1338F} {139, 1338}

\bibitem[\protect\citeauthoryear{{Gaia Collaboration} et~al.,}{{Gaia Collaboration} et~al.}{2016}]{2016A&A...595A...1G}
{Gaia Collaboration} et~al., 2016, \mn@doi [\aap] {10.1051/0004-6361/201629272}, \href {https://ui.adsabs.harvard.edu/abs/2016A&A...595A...1G} {595, A1}

\bibitem[\protect\citeauthoryear{{Gaia Collaboration} et~al.,}{{Gaia Collaboration} et~al.}{2018}]{2018A&A...616A...1G}
{Gaia Collaboration} et~al., 2018, \mn@doi [\aap] {10.1051/0004-6361/201833051}, \href {https://ui.adsabs.harvard.edu/abs/2018A&A...616A...1G} {616, A1}

\bibitem[\protect\citeauthoryear{{Gaia Collaboration} et~al.,}{{Gaia Collaboration} et~al.}{2023}]{2023A&A...674A...1G}
{Gaia Collaboration} et~al., 2023, \mn@doi [\aap] {10.1051/0004-6361/202243940}, \href {https://ui.adsabs.harvard.edu/abs/2023A&A...674A...1G} {674, A1}

\bibitem[\protect\citeauthoryear{{Goodman} et~al.,}{{Goodman} et~al.}{2014}]{2014ApJ...797...53G}
{Goodman} A.~A.,  et~al., 2014, \mn@doi [\apj] {10.1088/0004-637X/797/1/53}, \href {https://ui.adsabs.harvard.edu/abs/2014ApJ...797...53G} {797, 53}

\bibitem[\protect\citeauthoryear{{Goodwin} \& {Bastian}}{{Goodwin} \& {Bastian}}{2006}]{2006MNRAS.373..752G}
{Goodwin} S.~P.,  {Bastian} N.,  2006, \mn@doi [\mnras] {10.1111/j.1365-2966.2006.11078.x}, \href {https://ui.adsabs.harvard.edu/abs/2006MNRAS.373..752G} {373, 752}

\bibitem[\protect\citeauthoryear{{Gro{\ss}schedl}, {Alves}, {Meingast}  \& {Herbst-Kiss}}{{Gro{\ss}schedl} et~al.}{2021}]{2021A&A...647A..91G}
{Gro{\ss}schedl} J.~E.,  {Alves} J.,  {Meingast} S.,   {Herbst-Kiss} G.,  2021, \mn@doi [\aap] {10.1051/0004-6361/202038913}, \href {https://ui.adsabs.harvard.edu/abs/2021A&A...647A..91G} {647, A91}

\bibitem[\protect\citeauthoryear{{G{\"u}del}}{{G{\"u}del}}{2004}]{2004A&ARv..12...71G}
{G{\"u}del} M.,  2004, \mn@doi [\aapr] {10.1007/s00159-004-0023-2}, \href {https://ui.adsabs.harvard.edu/abs/2004A&ARv..12...71G} {12, 71}

\bibitem[\protect\citeauthoryear{{Ha}, {Li}, {Kounkel}, {Xu}, {Li}  \& {Zheng}}{{Ha} et~al.}{2022}]{2022ApJ...934....7H}
{Ha} T.,  {Li} Y.,  {Kounkel} M.,  {Xu} S.,  {Li} H.,   {Zheng} Y.,  2022, \mn@doi [\apj] {10.3847/1538-4357/ac76bf}, \href {https://ui.adsabs.harvard.edu/abs/2022ApJ...934....7H} {934, 7}

\bibitem[\protect\citeauthoryear{{Heyer}, {Krawczyk}, {Duval}  \& {Jackson}}{{Heyer} et~al.}{2009}]{2009ApJ...699.1092H}
{Heyer} M.,  {Krawczyk} C.,  {Duval} J.,   {Jackson} J.~M.,  2009, \mn@doi [\apj] {10.1088/0004-637X/699/2/1092}, \href {https://ui.adsabs.harvard.edu/abs/2009ApJ...699.1092H} {699, 1092}

\bibitem[\protect\citeauthoryear{{Kawamura} et~al.,}{{Kawamura} et~al.}{2009}]{2009ApJS..184....1K}
{Kawamura} A.,  et~al., 2009, \mn@doi [\apjs] {10.1088/0067-0049/184/1/1}, \href {https://ui.adsabs.harvard.edu/abs/2009ApJS..184....1K} {184, 1}

\bibitem[\protect\citeauthoryear{{Koenig}, {Allen}, {Gutermuth}, {Hora}, {Brunt}  \& {Muzerolle}}{{Koenig} et~al.}{2008}]{2008ApJ...688.1142K}
{Koenig} X.~P.,  {Allen} L.~E.,  {Gutermuth} R.~A.,  {Hora} J.~L.,  {Brunt} C.~M.,   {Muzerolle} J.,  2008, \mn@doi [\apj] {10.1086/592322}, \href {https://ui.adsabs.harvard.edu/abs/2008ApJ...688.1142K} {688, 1142}

\bibitem[\protect\citeauthoryear{{Kounkel}, {Covey}  \& {Stassun}}{{Kounkel} et~al.}{2020}]{2020AJ....160..279K}
{Kounkel} M.,  {Covey} K.,   {Stassun} K.~G.,  2020, \mn@doi [\aj] {10.3847/1538-3881/abc0e6}, \href {https://ui.adsabs.harvard.edu/abs/2020AJ....160..279K} {160, 279}

\bibitem[\protect\citeauthoryear{{Krause}, {Fierlinger}, {Diehl}, {Burkert}, {Voss}  \& {Ziegler}}{{Krause} et~al.}{2013}]{2013A&A...550A..49K}
{Krause} M.,  {Fierlinger} K.,  {Diehl} R.,  {Burkert} A.,  {Voss} R.,   {Ziegler} U.,  2013, \mn@doi [\aap] {10.1051/0004-6361/201220060}, \href {https://ui.adsabs.harvard.edu/abs/2013A&A...550A..49K} {550, A49}

\bibitem[\protect\citeauthoryear{{Krause}, {Charbonnel}, {Bastian}  \& {Diehl}}{{Krause} et~al.}{2016}]{2016A&A...587A..53K}
{Krause} M. G.~H.,  {Charbonnel} C.,  {Bastian} N.,   {Diehl} R.,  2016, \mn@doi [\aap] {10.1051/0004-6361/201526685}, \href {https://ui.adsabs.harvard.edu/abs/2016A&A...587A..53K} {587, A53}

\bibitem[\protect\citeauthoryear{{Kruijssen} et~al.,}{{Kruijssen} et~al.}{2019}]{2019Natur.569..519K}
{Kruijssen} J.~M.~D.,  et~al., 2019, \mn@doi [\nat] {10.1038/s41586-019-1194-3}, \href {https://ui.adsabs.harvard.edu/abs/2019Natur.569..519K} {569, 519}

\bibitem[\protect\citeauthoryear{{Krumholz} et~al.,}{{Krumholz} et~al.}{2014}]{2014prpl.conf..243K}
{Krumholz} M.~R.,  et~al., 2014, in {Beuther} H.,  {Klessen} R.~S.,  {Dullemond} C.~P.,   {Henning} T.,  eds, Protostars and Planets VI. p.~243 (\mn@eprint {arXiv} {1401.2473}), \mn@doi{10.2458/azu_uapress_9780816531240-ch011}

\bibitem[\protect\citeauthoryear{{Kuhn} et~al.,}{{Kuhn} et~al.}{2021}]{2021A&A...651L..10K}
{Kuhn} M.~A.,  et~al., 2021, \mn@doi [\aap] {10.1051/0004-6361/202141198}, \href {https://ui.adsabs.harvard.edu/abs/2021A&A...651L..10K} {651, L10}

\bibitem[\protect\citeauthoryear{{Lada} \& {Lada}}{{Lada} \& {Lada}}{2003}]{2003ARA&A..41...57L}
{Lada} C.~J.,  {Lada} E.~A.,  2003, \mn@doi [\araa] {10.1146/annurev.astro.41.011802.094844}, \href {https://ui.adsabs.harvard.edu/abs/2003ARA&A..41...57L} {41, 57}

\bibitem[\protect\citeauthoryear{{Lada}, {Strom}  \& {Myers}}{{Lada} et~al.}{1993}]{1993prpl.conf..245L}
{Lada} E.~A.,  {Strom} K.~M.,   {Myers} P.~C.,  1993, in {Levy} E.~H.,  {Lunine} J.~I.,  eds, Protostars and Planets III. p.~245

\bibitem[\protect\citeauthoryear{{Lallement}, {Babusiaux}, {Vergely}, {Katz}, {Arenou}, {Valette}, {Hottier}  \& {Capitanio}}{{Lallement} et~al.}{2019}]{2019A&A...625A.135L}
{Lallement} R.,  {Babusiaux} C.,  {Vergely} J.~L.,  {Katz} D.,  {Arenou} F.,  {Valette} B.,  {Hottier} C.,   {Capitanio} L.,  2019, \mn@doi [\aap] {10.1051/0004-6361/201834695}, \href {https://ui.adsabs.harvard.edu/abs/2019A&A...625A.135L} {625, A135}

\bibitem[\protect\citeauthoryear{{Larson}}{{Larson}}{1981}]{1981MNRAS.194..809L}
{Larson} R.~B.,  1981, \mn@doi [\mnras] {10.1093/mnras/194.4.809}, \href {https://ui.adsabs.harvard.edu/abs/1981MNRAS.194..809L} {194, 809}

\bibitem[\protect\citeauthoryear{{Larson}}{{Larson}}{1994}]{1994LNP...439...13L}
{Larson} R.~B.,  1994, in {Wilson} T.~L.,  {Johnston} K.~J.,  eds, , Vol.~439, The Structure and Content of Molecular Clouds.
p.~13, \mn@doi{10.1007/3540586210_2}

\bibitem[\protect\citeauthoryear{{Larson}}{{Larson}}{1995}]{1995MNRAS.272..213L}
{Larson} R.~B.,  1995, \mn@doi [\mnras] {10.1093/mnras/272.1.213}, \href {https://ui.adsabs.harvard.edu/abs/1995MNRAS.272..213L} {272, 213}

\bibitem[\protect\citeauthoryear{{Li} \& {Chen}}{{Li} \& {Chen}}{2022}]{2022arXiv220503218L}
{Li} G.-X.,  {Chen} B.-Q.,  2022, arXiv e-prints, \href {https://ui.adsabs.harvard.edu/abs/2022arXiv220503218L} {p. arXiv:2205.03218}

\bibitem[\protect\citeauthoryear{{Li}, {Wyrowski}, {Menten}  \& {Belloche}}{{Li} et~al.}{2013}]{2013A&A...559A..34L}
{Li} G.-X.,  {Wyrowski} F.,  {Menten} K.,   {Belloche} A.,  2013, \mn@doi [\aap] {10.1051/0004-6361/201322411}, \href {https://ui.adsabs.harvard.edu/abs/2013A&A...559A..34L} {559, A34}

\bibitem[\protect\citeauthoryear{{Li}, {Goodman}, {Sridharan}, {Houde}, {Li}, {Novak}  \& {Tang}}{{Li} et~al.}{2014}]{2014prpl.conf..101L}
{Li} H.~B.,  {Goodman} A.,  {Sridharan} T.~K.,  {Houde} M.,  {Li} Z.~Y.,  {Novak} G.,   {Tang} K.~S.,  2014, in {Beuther} H.,  {Klessen} R.~S.,  {Dullemond} C.~P.,   {Henning} T.,  eds, Protostars and Planets VI. p.~101 (\mn@eprint {arXiv} {1404.2024}), \mn@doi{10.2458/azu_uapress_9780816531240-ch005}

\bibitem[\protect\citeauthoryear{Lindsay}{Lindsay}{1995}]{lindsay1995mixture}
Lindsay B.~G.,  1995.

\bibitem[\protect\citeauthoryear{{Mac Low} \& {Klessen}}{{Mac Low} \& {Klessen}}{2004}]{2004RvMP...76..125M}
{Mac Low} M.-M.,  {Klessen} R.~S.,  2004, \mn@doi [Reviews of Modern Physics] {10.1103/RevModPhys.76.125}, \href {https://ui.adsabs.harvard.edu/abs/2004RvMP...76..125M} {76, 125}

\bibitem[\protect\citeauthoryear{{Marton}, {T{\'o}th}, {Paladini}, {Kun}, {Zahorecz}, {McGehee}  \& {Kiss}}{{Marton} et~al.}{2016}]{2016MNRAS.458.3479M}
{Marton} G.,  {T{\'o}th} L.~V.,  {Paladini} R.,  {Kun} M.,  {Zahorecz} S.,  {McGehee} P.,   {Kiss} C.,  2016, \mn@doi [\mnras] {10.1093/mnras/stw398}, \href {https://ui.adsabs.harvard.edu/abs/2016MNRAS.458.3479M} {458, 3479}

\bibitem[\protect\citeauthoryear{{McKee} \& {Ostriker}}{{McKee} \& {Ostriker}}{2007}]{2007ARA&A..45..565M}
{McKee} C.~F.,  {Ostriker} E.~C.,  2007, \mn@doi [\araa] {10.1146/annurev.astro.45.051806.110602}, \href {https://ui.adsabs.harvard.edu/abs/2007ARA&A..45..565M} {45, 565}

\bibitem[\protect\citeauthoryear{{Megeath} et~al.,}{{Megeath} et~al.}{2004}]{2004ApJS..154..367M}
{Megeath} S.~T.,  et~al., 2004, \mn@doi [\apjs] {10.1086/422823}, \href {https://ui.adsabs.harvard.edu/abs/2004ApJS..154..367M} {154, 367}

\bibitem[\protect\citeauthoryear{{Miville-Desch{\^e}nes}, {Murray}  \& {Lee}}{{Miville-Desch{\^e}nes} et~al.}{2017}]{2017ApJ...834...57M}
{Miville-Desch{\^e}nes} M.-A.,  {Murray} N.,   {Lee} E.~J.,  2017, \mn@doi [\apj] {10.3847/1538-4357/834/1/57}, \href {https://ui.adsabs.harvard.edu/abs/2017ApJ...834...57M} {834, 57}

\bibitem[\protect\citeauthoryear{{Pang}, {Li}, {Tang}, {Pasquato}  \& {Kouwenhoven}}{{Pang} et~al.}{2020}]{2020ApJ...900L...4P}
{Pang} X.,  {Li} Y.,  {Tang} S.-Y.,  {Pasquato} M.,   {Kouwenhoven} M.~B.~N.,  2020, \mn@doi [\apjl] {10.3847/2041-8213/abad28}, \href {https://ui.adsabs.harvard.edu/abs/2020ApJ...900L...4P} {900, L4}

\bibitem[\protect\citeauthoryear{Pedregosa et~al.,}{Pedregosa et~al.}{2011}]{pedregosa2011scikit}
Pedregosa F.,  et~al., 2011, the Journal of machine Learning research, 12, 2825

\bibitem[\protect\citeauthoryear{{Planck Collaboration} et~al.,}{{Planck Collaboration} et~al.}{2014}]{2014A&A...571A...1P}
{Planck Collaboration} et~al., 2014, \mn@doi [\aap] {10.1051/0004-6361/201321529}, \href {https://ui.adsabs.harvard.edu/abs/2014A&A...571A...1P} {571, A1}

\bibitem[\protect\citeauthoryear{{Ragan}, {Henning}, {Tackenberg}, {Beuther}, {Johnston}, {Kainulainen}  \& {Linz}}{{Ragan} et~al.}{2014}]{2014A&A...568A..73R}
{Ragan} S.~E.,  {Henning} T.,  {Tackenberg} J.,  {Beuther} H.,  {Johnston} K.~G.,  {Kainulainen} J.,   {Linz} H.,  2014, \mn@doi [\aap] {10.1051/0004-6361/201423401}, \href {https://ui.adsabs.harvard.edu/abs/2014A&A...568A..73R} {568, A73}

\bibitem[\protect\citeauthoryear{{Ribas}, {Mer{\'\i}n}, {Bouy}  \& {Maud}}{{Ribas} et~al.}{2014}]{2014A&A...561A..54R}
{Ribas} {\'A}.,  {Mer{\'\i}n} B.,  {Bouy} H.,   {Maud} L.~T.,  2014, \mn@doi [\aap] {10.1051/0004-6361/201322597}, \href {https://ui.adsabs.harvard.edu/abs/2014A&A...561A..54R} {561, A54}

\bibitem[\protect\citeauthoryear{{Ribas}, {Bouy}  \& {Mer{\'\i}n}}{{Ribas} et~al.}{2015}]{2015A&A...576A..52R}
{Ribas} {\'A}.,  {Bouy} H.,   {Mer{\'\i}n} B.,  2015, \mn@doi [\aap] {10.1051/0004-6361/201424846}, \href {https://ui.adsabs.harvard.edu/abs/2015A&A...576A..52R} {576, A52}

\bibitem[\protect\citeauthoryear{{Rodriguez}, {Bessell}, {Zuckerman}  \& {Kastner}}{{Rodriguez} et~al.}{2011}]{2011ApJ...727...62R}
{Rodriguez} D.~R.,  {Bessell} M.~S.,  {Zuckerman} B.,   {Kastner} J.~H.,  2011, \mn@doi [\apj] {10.1088/0004-637X/727/2/62}, \href {https://ui.adsabs.harvard.edu/abs/2011ApJ...727...62R} {727, 62}

\bibitem[\protect\citeauthoryear{{Rodriguez}, {Zuckerman}, {Kastner}, {Bessell}, {Faherty}  \& {Murphy}}{{Rodriguez} et~al.}{2013}]{2013ApJ...774..101R}
{Rodriguez} D.~R.,  {Zuckerman} B.,  {Kastner} J.~H.,  {Bessell} M.~S.,  {Faherty} J.~K.,   {Murphy} S.~J.,  2013, \mn@doi [\apj] {10.1088/0004-637X/774/2/101}, \href {https://ui.adsabs.harvard.edu/abs/2013ApJ...774..101R} {774, 101}

\bibitem[\protect\citeauthoryear{{Rosolowsky}, {Pineda}, {Kauffmann}  \& {Goodman}}{{Rosolowsky} et~al.}{2008}]{2008ApJ...679.1338R}
{Rosolowsky} E.~W.,  {Pineda} J.~E.,  {Kauffmann} J.,   {Goodman} A.~A.,  2008, \mn@doi [\apj] {10.1086/587685}, \href {https://ui.adsabs.harvard.edu/abs/2008ApJ...679.1338R} {679, 1338}

\bibitem[\protect\citeauthoryear{{Vazquez-Semadeni}, {Ostriker}, {Passot}, {Gammie}  \& {Stone}}{{Vazquez-Semadeni} et~al.}{2000}]{2000prpl.conf....3V}
{Vazquez-Semadeni} E.,  {Ostriker} E.~C.,  {Passot} T.,  {Gammie} C.~F.,   {Stone} J.~M.,  2000, in {Mannings} V.,  {Boss} A.~P.,   {Russell} S.~S.,  eds, Protostars and Planets IV. p.~3 (\mn@eprint {arXiv} {astro-ph/9903066})

\bibitem[\protect\citeauthoryear{{Wilking} \& {Lada}}{{Wilking} \& {Lada}}{1983}]{1983ApJ...274..698W}
{Wilking} B.~A.,  {Lada} C.~J.,  1983, \mn@doi [\apj] {10.1086/161482}, \href {https://ui.adsabs.harvard.edu/abs/1983ApJ...274..698W} {274, 698}

\bibitem[\protect\citeauthoryear{{Wyatt}}{{Wyatt}}{2008}]{2008ARA&A..46..339W}
{Wyatt} M.~C.,  2008, \mn@doi [\araa] {10.1146/annurev.astro.45.051806.110525}, \href {https://ui.adsabs.harvard.edu/abs/2008ARA&A..46..339W} {46, 339}

\bibitem[\protect\citeauthoryear{{Zari}, {Hashemi}, {Brown}, {Jardine}  \& {de Zeeuw}}{{Zari} et~al.}{2018}]{2018A&A...620A.172Z}
{Zari} E.,  {Hashemi} H.,  {Brown} A.~G.~A.,  {Jardine} K.,   {de Zeeuw} P.~T.,  2018, \mn@doi [\aap] {10.1051/0004-6361/201834150}, \href {https://ui.adsabs.harvard.edu/abs/2018A&A...620A.172Z} {620, A172}

\bibitem[\protect\citeauthoryear{{Zhang}}{{Zhang}}{2023}]{2023ApJS..265...59Z}
{Zhang} M.,  2023, \mn@doi [\apjs] {10.3847/1538-4365/acc1e8}, \href {https://ui.adsabs.harvard.edu/abs/2023ApJS..265...59Z} {265, 59}

\bibitem[\protect\citeauthoryear{{Zhou}, {Li}  \& {Chen}}{{Zhou} et~al.}{2022}]{2022MNRAS.513..638Z}
{Zhou} J.-X.,  {Li} G.-X.,   {Chen} B.-Q.,  2022, \mn@doi [\mnras] {10.1093/mnras/stac900}, \href {https://ui.adsabs.harvard.edu/abs/2022MNRAS.513..638Z} {513, 638}

\bibitem[\protect\citeauthoryear{{Zinnecker}, {McCaughrean}  \& {Wilking}}{{Zinnecker} et~al.}{1993}]{1993prpl.conf..429Z}
{Zinnecker} H.,  {McCaughrean} M.~J.,   {Wilking} B.~A.,  1993, in {Levy} E.~H.,  {Lunine} J.~I.,  eds, Protostars and Planets III. p.~429

\bibitem[\protect\citeauthoryear{{Zucker} et~al.,}{{Zucker} et~al.}{2022}]{2022Natur.601..334Z}
{Zucker} C.,  et~al., 2022, \mn@doi [\nat] {10.1038/s41586-021-04286-5}, \href {https://ui.adsabs.harvard.edu/abs/2022Natur.601..334Z} {601, 334}

\makeatother
\end{thebibliography}

\appendix
\section{Finding the correct velocity component from CO data}
\label{appendixa}
\begin{figure*}
\begin{center}
\includegraphics[scale=0.8]{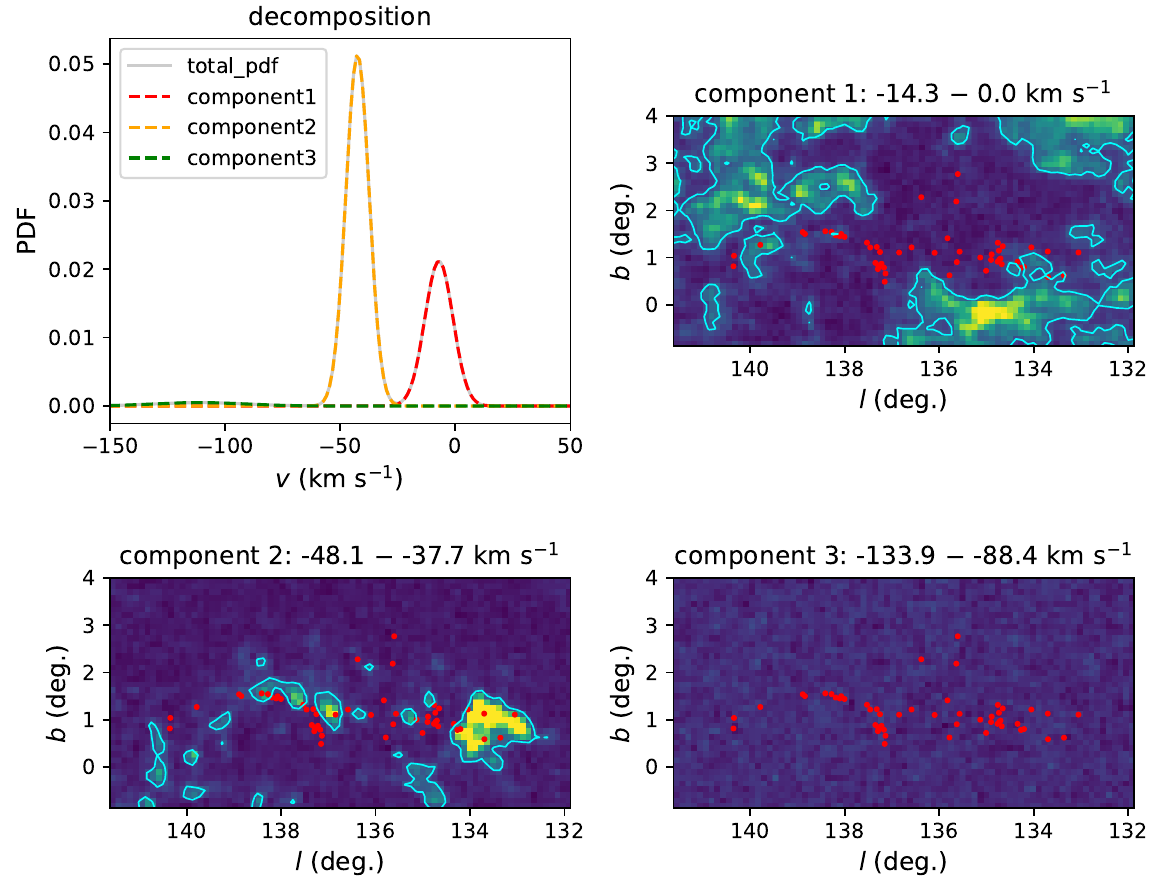}
\caption{{\bf Components selection in the CO data towards a Type 1 direct association.} The upper left panel shows the PDF (Probability density function) distribution for the best-fitting CO line profile integrated over the association region. The grey, red, orange, and green lines refer to the PDF of the total distribution and three components, respectively. The red dots in the rest three subplots show the location of member YSOs. The background is CO maps obtained by integrating over velocity range in the title. The cyan contours mark the region with CO emission larger than 2.21 K.}
\label{A_direct}
\end{center}
\end{figure*}

In the CO data, which is presented in the form of a cube in the position-position-velocity space, different clouds can appear as different emission peaks along the velocity axis.
Towards each association, our aim is to identify cloud counterparts from the multiple components located at
different velocity channels in CO data. First, according to the footprint of each
YSO association, we select a larger area of CO data and set the value of the
pixel without member YSOs as 0 at all velocity channels. The modified CO data
are integrated into $l\&b$ space and then used to produce the line
profiles along the velocity. We apply the \texttt{Gaussian.Mixture} algorithm
to the line profile and set the Gaussian components number from 1 to 3. Based on
the AIC criteria, the algorithm gives the best-fitting result with the smallest AIC value. As shown in the upper left panel in Fig. \ref{A_direct}, the PDF (Probability density function) of the best-fitting components is given. For each component,
we calculate the FWHM (Full Width  Half Maxima) range based on the velocity component and
integrate original CO data at $[v_{\rm lower}, v_{\rm upper}]$ to match the YSO association, where $v_{\rm lower}$ and $v_{\rm upper}$ mark the FWHM for the PDF of the velocity component. As
presented in three subplots in Fig. \ref{A_direct}, we
compare the distributions of member YSOs (red dots) with the morphology of the CO-emitting components (in cyan contours) integrated over different components. Cyan contours in three subplots mark the area with emission larger than 2.21 K, which is the maximum standard deviation in component 1, component 2, and component 3. From the upper right and lower left panel in Fig. \ref{A_direct},  strong emissions are found in the background
towards components 1 and 2 in cyan contours. By visual check, we take component 2 as the gas counterpart because the member YSOs share a similar morphology with the gas.
Through this process, we determine the gas content for each YSO association by visual check, as well as the types of the YSO associations.  

% \begin{figure*}
% \begin{center}
% \includegraphics[scale=0.6]{clean_asso0605.pdf}
% \caption{{\bf Components selection in CO data towards a \emph{clean} association Type.} The upper left panel shows the line profile of CO data integrated over the chosen region. The grey, red, orange, and green lines refer to the PDF (Probability distribution function) of the total distribution and three components, respectively. The red dots in the rest three subplots show the location of member YSOs and red contours mark the YSO distribution. The background is CO maps obtained by integrating over the three velocity components and yellow contours mark the region with CO emission. The range of the integrations is indicated as the titles of the subplots.}
% \label{A_clean}
% \end{center}
% \end{figure*}

\section{Examples of association classification}
\label{examples for classification}
After finding the associated CO velocity component for most YSO associations, we compare the location of member YSOs with both Planck dust emission and CO emission to classify the YSO associations. We plot the member YSOs on the dust map and integrated the CO map over the velocity range for the best-matched component.

\begin{figure*}
\begin{center}
\includegraphics[scale=0.7]{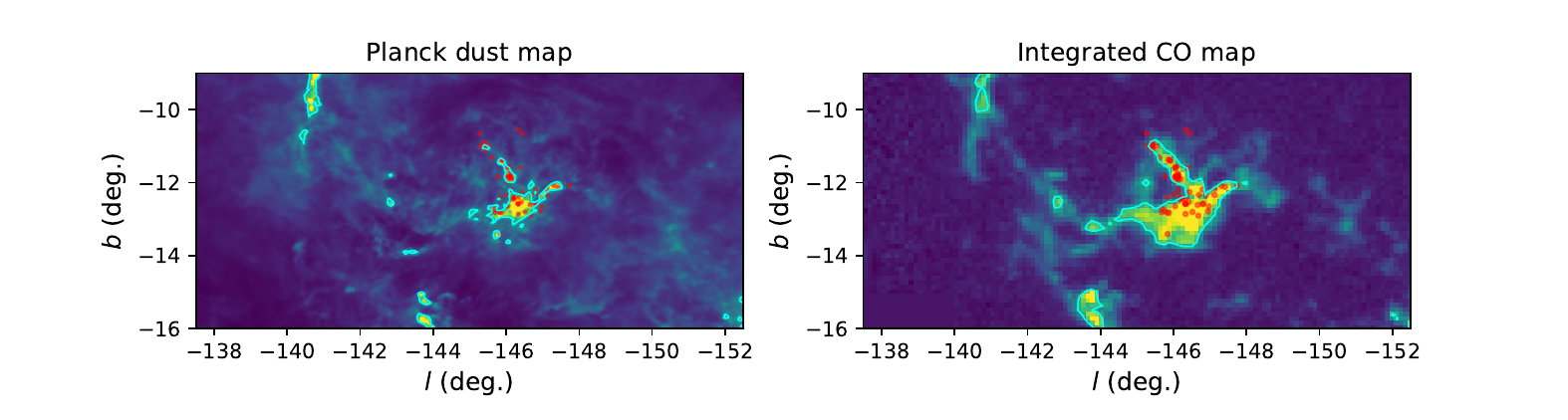}
\caption{{\bf Example for a Type 1 direct association.} The red dots in both subplots are member YSOs. The background in the left panel is the Planck 857 GHz map and the background in the right panel is the CO map integrated from 10.4 to 13 km s$^{-1}$. In both panels, the cyan contours mark the area with emission larger than 3$\sigma$. Almost all the YSOs are located inside the gas region.}
\label{example_1.1}
\end{center}
\end{figure*}

We show a Type 1 direct association example in Fig. \ref{example_1.1}. CO map shown in the right panel is integrated from 10.4 to 13 km s$^{-1}$. We use cyan contours marking the area with emission larger than 3 sigma in both the Planck map and the CO map. We compare the red dots (member YSOs) with the cyan contours. In the right panel, nearly all the member YSOs are located inside the gas region, making it a Type 1 direct association.

\begin{figure*}
\begin{center}
\includegraphics[scale=0.7]{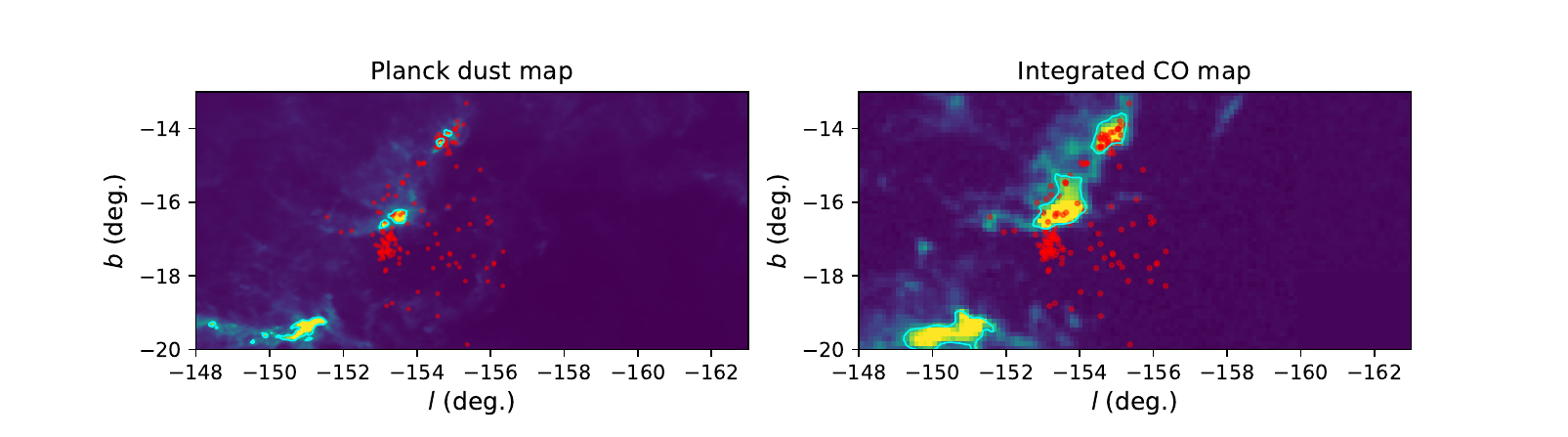}
\caption{{\bf Example foofr a Type 2 close association.} All the markers are the same as Fig. \ref{example_1.1}. In the right panel, the CO map is integrated from 7.8 to 10.4 km s$^{-1}$. The member YSOs are partially associated with the gas distribution.}
\label{example_1.2}
\end{center}
\end{figure*}

We show a Type 2 close association example in Fig. \ref{example_1.2}. CO map shown in the right panel is integrated from 7.8 to 10.4 km s$^{-1}$. In both panels, the red dots are partially outside the cyan gas contours. Being partially associated with gas makes them a Type 2 close association.

\begin{figure*}
\begin{center}
\includegraphics[scale=0.7]{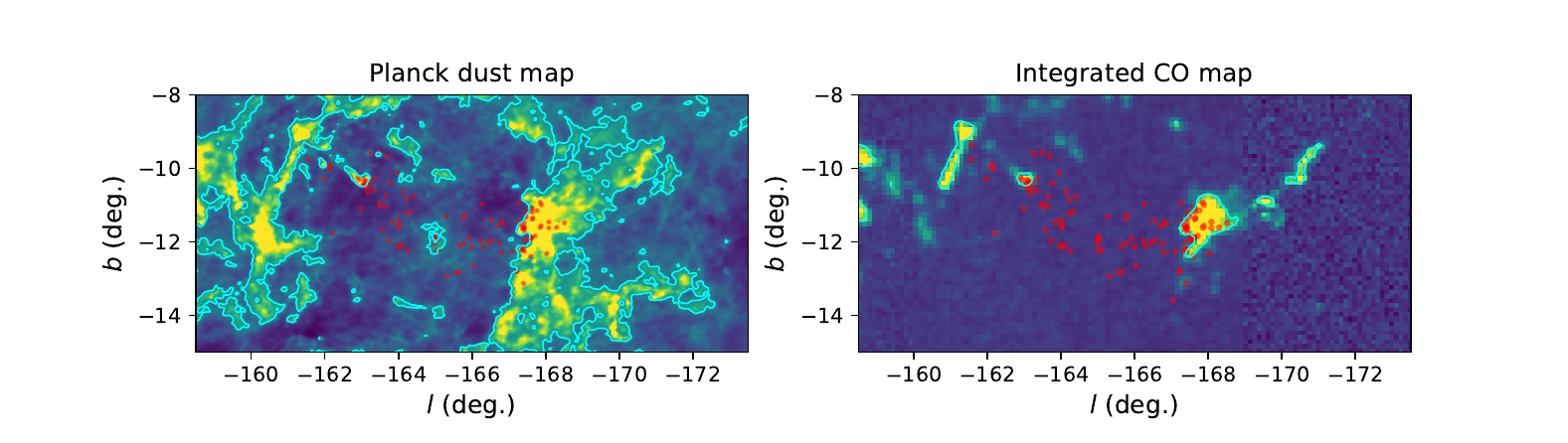}
\caption{{\bf Example of a Type 3 bubble association.} All the markers are the same as Fig. \ref{example_1.1}. In the right panel, the CO map is integrated from 7.8 to 11.7 km s$^{-1}$. The member YSOs partially match the bubble-like gas distribution.}
\label{example_1.3}
\end{center}
\end{figure*}

We show a Type 3 bubble association example in Fig. \ref{example_1.3}. The CO map shown in the right panel is integrated from 7.8 to 11.7 km s$^{-1}$. The red YSO contours partially coincide with the gas contours. From the Planck dust map in the left panel, it shows a clear bubble-like gas structure. Being partially associated with a bubble structure makes it a Type 3 bubble association.

\begin{figure*}
\begin{center}
\includegraphics[scale=0.7]{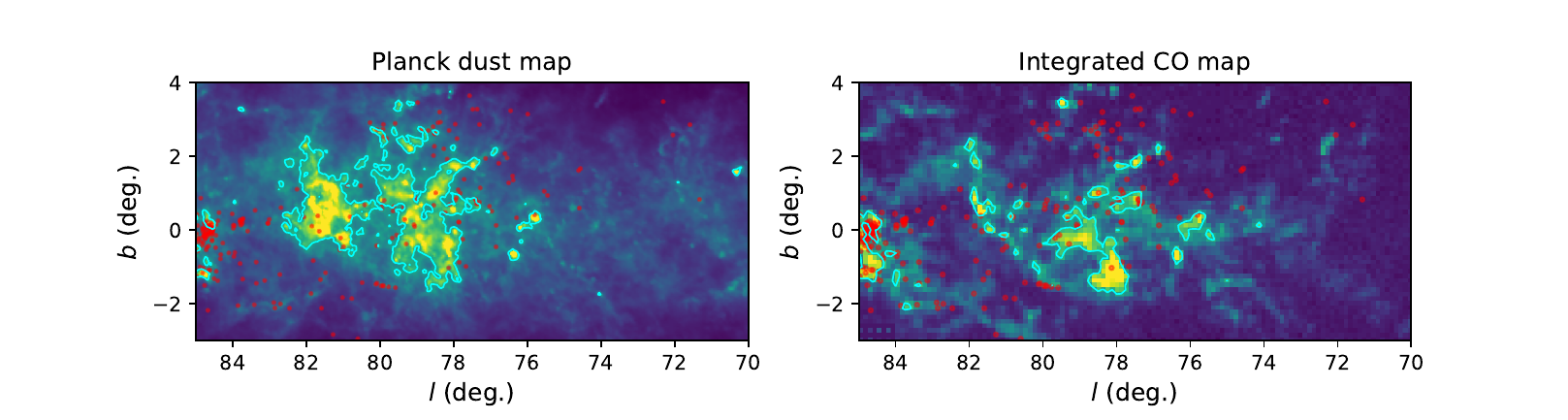}
\caption{{\bf Example of a Type 4 complex association.} All the markers are the same as Fig. \ref{example_1.1}.In the right panel, CO map is integrated from $-5.2$ to 3.9 km s$^{-1}$. Part of the member YSOs are associated with gas and part are only associated with very diffuse gas structure.}
\label{example_3}
\end{center}
\end{figure*}

We show a Type 4 complex association example in Fig. \ref{example_3}. The CO map shown in the right panel is integrated from $-5.2$ to 3.9 km s$^{-1}$. From both the Planck map and the  CO map, part of the member YSOs are located inside cyan gas contours, and part of them are only associated with diffuse gas structures. This kind of complex relation with surrounding gas makes them Type 4 complex associations. They are usually the branch structures. They can be good samples for large association research.

\begin{figure*}
\begin{center}
\includegraphics[scale=0.7]{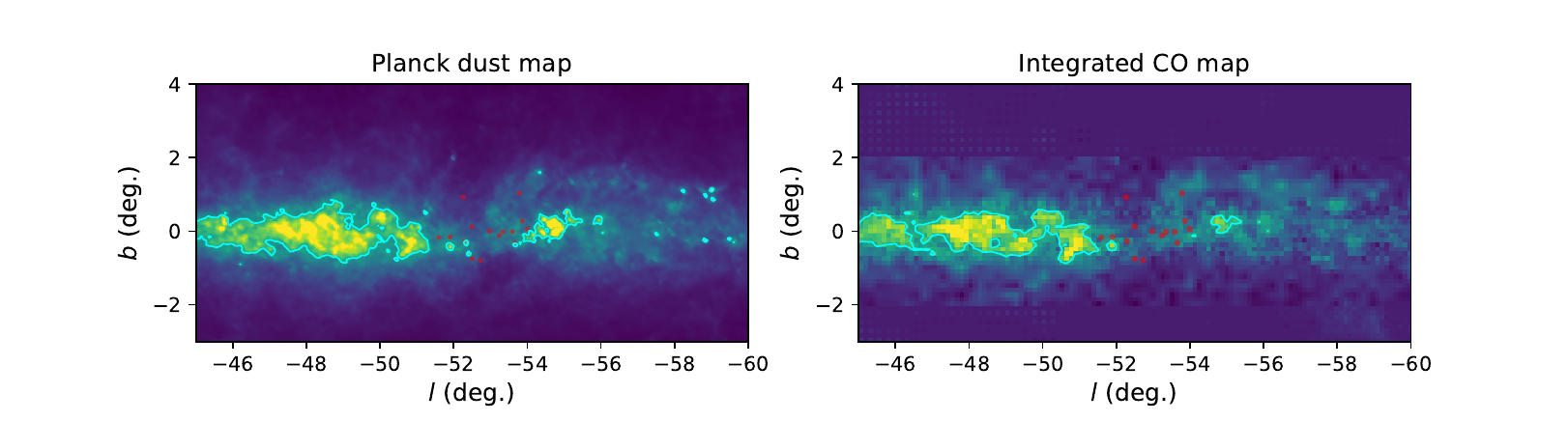}
\caption{{\bf Example of a Type 5 diffuse association.} All the markers are the same as Fig. \ref{example_1.1}. In the right panel, the CO map is integrated over the whole velocity range. The YSO distribution matches a very diffuse gas distribution. }
\label{example_2.1}
\end{center}
\end{figure*}

We show a Type 5 diffuse association example in Fig. \ref{example_2.1}. The CO map shown in the right panel is integrated over the whole range. The red dots are not associated with the strong emission but diffuse gas morphology can be seen. For gas-poor associations, we also check them with a 3D dust map from \citet{2019MNRAS.483.4277C}.

\begin{figure*}
\begin{center}
\includegraphics[scale=0.7]{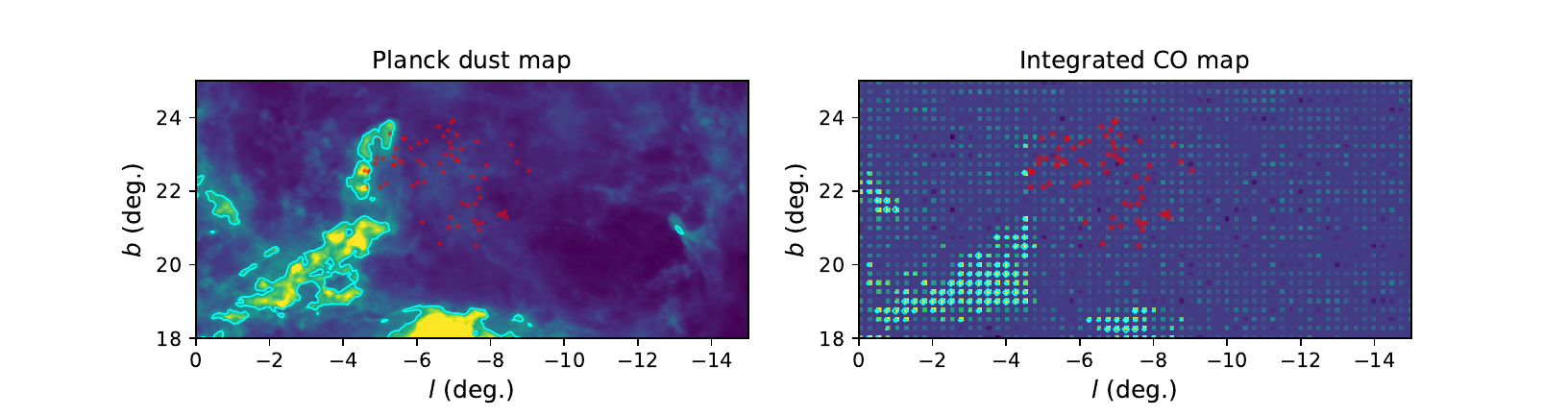}
\caption{{\bf Example for a Type 6 clean association.} All the markers are the same as Fig. \ref{example_1.1}. In the right panel, the CO map is integrated over the whole velocity range. The YSO distribution seldom matches the gas distribution.}
\label{example_2.2}
\end{center}
\end{figure*}

We show a Type 6 clean association example in Fig. \ref{example_2.2}. The CO map shown in the right panel is integrated over the whole range. In both the Planck map and the CO map, the red dots are not associated with gas emission. It's been checked with a 3D dust map from \citet{2019MNRAS.483.4277C}.

\clearpage
\onecolumn

\end{document}